\documentclass[a4paper]{article}
\usepackage{RRA4}
\RRNo{6513}
\usepackage{hyperref}
\usepackage{amssymb}
\usepackage{multirow}
\RRdate{April 2008}

\RRauthor{
Van Hoa NGUYEN, Dominique LAVENIER
}
\authorhead{Nguyen \& Lavenier}
\RRtitle{Parall\'{e}lisation \`{a} grain fin de la recherche de similarit\'{e}s entre s\'{e}quences prot\'{e}iques}
\RRetitle{Fine-grained parallelization of similarity search between protein sequences}
\titlehead{Fine-grained parallelization }
\RRnote{}
\RRnote{}

\RRresume{Ce rapporte pr\'{e}sente l'impl\'{e}mentation d'un agorithme
de   comparaison  de  s\'{e}quences   prot\'{e}iques  sp\'{e}cialement
con\c{c}u  pour que  les parties  les plus  co\^{u}teuses en  temps de
calcul puissent s'ex\'{e}cuter  en parall\`{e}le sur des architectures
\`{a} jeu  d'instruction SSE,  des architectures multi-coeurs,  ou des
cartes   graphiques  de   derni\`{e}res   g\'{e}n\'{e}rations.   Trois
programmes ont \'{e}t\'{e} d\'{e}velop\-p\'{e}s~: PLAST-P, TPLAST-N et
PLAST-X.  Ils  g\'{e}n\`{e}rent des r\'{e}sultats  \'{e}quivalents aux
programmes  de  la  famille   BLAST  (BLAST-P,  TBLAST-P  et  BLAST-X)
d\'{e}velo- pp\'{e}s au NCBI. Les facteurs d'acc\'{e}l\'{e}ration (par
rapport \`{a} BLAST) s'\'{e}che\-lonnent de 5 \`{a} 10.

}

\RRabstract{This  report  presents  the  implementation of  a  protein
sequence  comparison algorithm specifically  designed for  speeding up
time  consuming part on  parallel hardware  such as  SSE instructions,
multicore  architectures or graphic  boards.  Three programs have been
developed:  PLAST-P, TPLAST-N  and PLAST-X.   They  provide equivalent
results compared to the  NCBI BLAST family programs (BLAST-P, TBLAST-N
and BLAST-X) with a speed-up factor ranging from 5 to 10.

}

\RRmotcle{Parall\'{e}lisation, recherche de similarit\'{e}s, indexation, BLAST, GPU, SIMD}
\RRkeyword{Parallelization, similarity search, indexing, BLAST, GPU, SIMD}
\RRprojets{Symbiose}
\RRtheme{\THBio} 
\RCRennes 

\begin{document}
\makeRR   

\newpage
\tableofcontents
\newpage

\section{Introduction}

In genomic,  similarity search aims  to find local  alignments between
two  DNA or  protein sequences,  measured by  match, mismatch  and gap
scores. Its objective is to locate regions in DNA or protein sequences
having closed relationship.  A typical application, for example, is to
query a  bank with a  gene whose function  is unknown in order  to get
some clues for further investigation.

Recent biotechnology improvements in the  sequencing area have led to a
huge   increasing  in   the  size   of  genomic   databases.   Genbank
\cite{Benson07},  for  example, contains  more  than  193 billions  of
nucleotides (February  2008) and  its size is  multiplied by  a factor
ranging from 1.4 and 1.5 every year.

Several algorithms have been proposed  to find alignments.  One of the
first, known  as the Smith-Waternam  algorithm, has been  developed in
1981  \cite{SMITH}  to  detect  local  alignments.   It  uses  dynamic
programming techniques  and has a quadratic  complexity. Another one,
BLAST, developed  in 1990,  is currently the  reference in  the domain
\cite{Altschul90}\cite{Altschul97}.   This  algorithm  is based  on  a
powerful heuristic coming from the following observation: an alignment
includes at least a word of  W characters (called a hit) shares by the
two sequences. Thus, instead of  exploring a large search space, as it
is done  in the dynamic programming  technique, hits are  first used to
target  small   zones  of   high  similarity  before   computing  full
alignments.

Many  nucleotide  search  tools,  using  this  heuristics,  have  been
proposed  to  perform  hit  detection: MegaBLAST  \cite{Kent02},  BLAT
\cite{Williams02}  or   PartternHunter  \cite{Li04}  \cite{Ma02},  for
example, include this technique.

These programs (except BLAT) have  been  mainly  designed for scanning
genomic databases, that is, given an input sequence (a gene, or a protein),
find in  the database  the other genes  or other proteins  which share
common similarities. There  are also used in the  context of intensive
sequence comparison whose purpose is to compare two genomic databases.
In that  case, they are not used  in an optimal way,  leading room for
further algorithm improvement.

This report details the results obtained from three programs developed
at  IRISA:  PLAST-P,  TPLAST-N  and  PLAST-X. PLAST  stands  for  {\it
Parallel Local  Alignment Search Tools}  by comparison to  BLAST ({\it
Basic  Local Alignment Search  Tools}). They  perform comparison  of 2
genomic banks  at the protein level  by first indexing  each bank, and
then,  by making  successive  refinements to  compute alignments.   In
addition,  hits are found  using subset-seeds  \cite{Perterlongo07} to
optimize memory access.

As  in the  BLAST  family, PLAST-P  takes  as input  2 protein  banks,
TPLAST-N  takes as input  a protein  bank and  a nucleotide  bank, and
PLAST-X  takes  as  input  two  nucleotide banks.   In  the  two  last
programs, the  nucleotide banks are  translated into 6  reading frames
and   the   comparison  is   made   using   amino  acid   substitution
cost. Different  implementations have been designed  ranging from pure
sequential version  to highly  parallel version using  modern graphics
processing units (GPU).

Practically, these programs need to  have the input banks in the FASTA
format.  The alignments are  generated following the -m~8 BLAST output
format.   As these  programs  are  expected to  be  used in  intensive
comparison contexts,  this format is  well suited for  automating post
processing.  Table \ref{ExampleBLASTOutput}  shows an example of three
alignments in this format (-m~8 option). An alignment is summarized on
a single  line with  the following items:  contents ID of  sequence in
bank1 and  sequence in bank2, alignment length,  position of alignment
on the two sequences, E-value and bit score.

\begin{table}[h]
\centering \small
\begin{tabular}{|c|cccccccc|}
\hline          &qry. id   & sub. id  &alig. len.&q. start&q. end&s. start& s. end&bit score\\
\hline 1        &P93208  & gi0689 &249     &1     &247     &1     &243  &296.3 \\
\hline 2        &P62261  & gi0689 &236     &2     &236     &1     &233  &309.4 \\
\hline 3        &P30488  & gi2750 &93      &217   &305     &64    &155  &43.9  \\
\hline
\end{tabular}
\caption{\small \emph{Example of BLAST output with option -m~8.}}
\label{ExampleBLASTOutput}
\end{table}

The  remainder of this  report is  organized as  follows: In  the next
section, a brief overview of the BLAST-P algorithm is given. Section 3
presents  the PLAST-P  algorithm.   Section 4  details the  sequential
version  of PLAST-P.   Section  5 describes  the  parallel version  of
PLAST-P.  Section 6 and 7 report results of TPLAST-N and PLAST-X.

\newpage
\section{BLAST-P algorithm}

BLAST is  one of the most popular bioinformatics tools and  is used to
run millions of queries every day. A family of BLAST programs has been
developed, depending on the type  of input.  According to these input,
the following naming is done: BLAST-P  when the query is an amino acid
sequence and the database is  made of protein sequences; TBLAST-N when
the  query is  an amino  acid  sequence and  the database  is made  of
nucleotide sequences; BLAST-X when the query is a nucleotide  sequence 
and the database is make of amino acid sequences. This section briefly 
describes  the  principle  of the BLAST-P  algorithm. It will help for  
understanding the difference with the PLAST-P algorithm.

\subsection{Context}

Generally,  in database  search,  when given  a  protein sequence,  the
objective is to extract from  a database all the similar sequences, or
zones of  similarity. The biological  motivation of this  operation is
generally to  assign a function  to unknown  genes or proteins:  in a
cell, a protein adopts a specific  3D shape related to its sequence of
amino  acids.  This  3D  structure is  important  because the  protein
function and  its interaction with  other molecules are  determined by
this  structure.  Two proteins  with  nearly  identical  3D shape  are
assumed to have similar functionalities. Thus, finding similar
sequences  is correlated  to find similar  structures and  then  to similar
functionalities.

Furthermore, today (April 2008) there are about 760 genomes being completely
sequenced,  and there  are more  than 3600  other  ongoing sequencing projects
\cite{genonesonline}. Now,  entire genomes are considered  as the main
pivot in  many bioinformatics research projects.  Genes  are no longer
considered  separately. The  genes of  a same  family  across multiple
species, for example,  may represent the abstract object  one wants to
study.  Thus,  instead of manipulating sequences with  a few thousands
of characters (genes), the needs have exploded to processing sequences
of a few millions of characters, called intensive computation.

Database  search  has to  explore  large  genomic  banks (hundreds  of
billions of  nucleotides), while intensive computations  works on much
less data. Suppose, that comparing two strings of respectively $L_{1}$
and   $L_{2}$   characters  has   a   complexity   of  $L_{1}   \times
L_{2}$.  Searching  a  sequence  of  size $10^{4}$  over  a  $10^{10}$
character database has the same complexity as comparing two strings of
size $10^{7}$.

In the first case, the problem  is I/O bounded: the search time mainly
depends  on the  bandwidth  capacity  of the  system  to transfer  the
database from the disk to the  CPU. In the second case, the problem is
compute  bounded   since  the  search  time  mainly   depends  on  the
computational  resources   available  for  computing  alignments.  The
BLAST-P  family programs  have  been primarily  designed for  scanning
large databases. Their use in intensive comparison context is possible,
but don't provide optimal performances.

\subsection{Generic algorithm}

This  section gives  an overview  of  the BLAST  family algorithm  for
searching proteins.  We will  focus on BLAST-P,  but the  TBLAST-N and
BLAST-X  extensions are governed  by similar  procedures and  won't be
detailed.

BLAST-P can be described by the following algorithm when comparing two
protein banks referred as bank1 and bank2.

\begin{center}
\small{
\begin{tabular}{ll}
\hline \multicolumn{2}{c}{BLAST-P Algorithm}                                                    \\
\hline
               for all sequences in bank1                                                   \\
\hspace{0.40cm}   make index                                                     &\#stage 0 \\
\hspace{0.40cm}   for all sequences in bank2  \textit{~~~~~~~~// scan of bank2}  &          \\
\hspace{0.80cm}       compute double hits from index                             &\#stage 1 \\
\hspace{0.80cm}       for all double hits                                        &          \\
\hspace{1.20cm}           compute ungapped alignment                             &\#stage 2 \\
\hspace{1.20cm}           if score $\geq S_{1}$                                  &          \\
\hspace{1.60cm}              compute gapped alignment                            &\#stage 3 \\
\hspace{1.60cm}              if score $\geq S_{2}$                               &          \\
\hspace{2.00cm}                 trace-back \& display alignment                   & \#stage 4\\
\hline
\end{tabular}}
\end{center}

In  the context  of intensive  genomic computation  between  bank1 and
bank2,  bank2 is  scanned N  times (N  is the  number of  sequences of
bank1).   From  a  computational  point  view, this  approach  is  not
efficient since  bank2 need to be  read a potentially  large number of
times. Actually, from version 2.2.10  to 2.2.13, the -B N option has
been added to search block  of multiple sequences.  This option allows
the  user  to  concatenate  all  sequences  in  bank1  into  a  single
one. Thus, the  argument of the -B option must be  equal to the number
of sequences in bank1. Processing multiple sequences in one run can be
much faster  than processing them separately because  bank2 is scanned
only one  time. From version  2.2.14 to 2.2.17,  BLAST-P automatically
concatenates several sequences (in  bank1) and  then compares  it with
bank2.   The  concatenated size  is  about  10,000  amino acids.   The
execution time is improved, but  for processing large database this is
still not sufficient.

BLAST-P can be  split into 5 different stages (from 0  to 4) which are
now described in the next sections.

\subsection{The 5 stages of the BLAST-P algorithm}

\subsubsection{Stage 0: Indexing.}

This first stage  computes an index structure for  each sequence of
bank1 using  words of  W characters \cite{Zhang98}.   To do  this, the
sequence  is  parsed  into  fixed-length overlapping  subsequences  of
size W.  For  example, if we suppose W = 3,  then, from the following
sequence NTHELYSLEISPQ, words  are: NTH, THE, HEL, ELY,  LYS, YSL, SLE,
and  so on.  These  words are  called  seeds since  they  are used  to
generated hits.

\begin{figure}[h]
  \centering
  \includegraphics[width=4.5in]{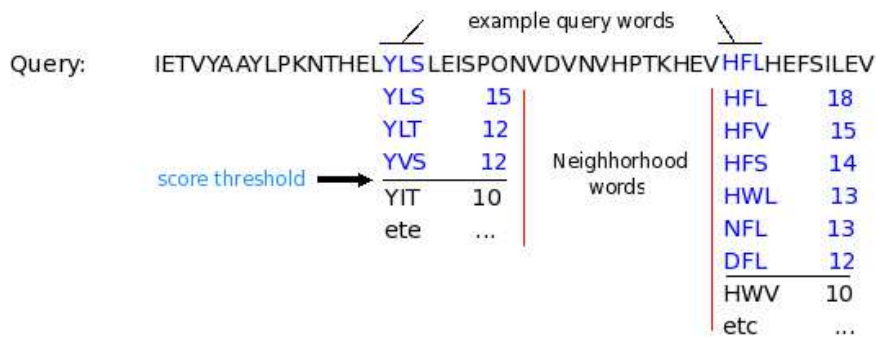}
  \caption{\small \emph{Example of neighborhood seeds with threshold equal to 11.}}\label{Fig:NeighborhoodWords}
\end{figure}

To augment the sensibility,  BLAST-P computes neighborhood seeds.  They
are identified by  computing a score between pairs  of words using
substitution  costs from a  substitution matrix.  Pairs having  a score
greater  than  a  predefined  threshold  value  T  are  considered  as
neighborhood seeds.  Figure \ref{Fig:NeighborhoodWords} illustrates the
identification of neighborhood  seeds tacking a threshold value  T = 11
(default value of BLAST-P).

More formally, two seeds $A_1, ... ,A_i$ and $B_1, ... B_i$ are considered as
neighborhood seeds if:

\[ \sum_{k=1}^i Sub(A_k,B_k) \geq T \]

\noindent
with $Sub(A,B)$ the cost for substituting A by B.

Once neigborhood seeds  have been identified, they are are stored into
a lookup table acting as an index structure.  Typically, seeds of 3
or 4 amino acids are taken (default value of BLAST-P is 3).

\subsubsection{Stage 1: Double hit computation}

When two  seeds coming  respectively  from the query sequence  and the
bank sequence match, the seed positions are labeled as a hit. It can be
seen as an anchoring point between the two sequences to locate zone of
potential alignment.

BLAST-P starts an alignment if two  hits are found within a small zone
with  no  gaps. To determine  these zones,  hits are  assigned  with a
diagonal number.  Hits  with identical  diagonal  numbers, and  closed
to each other, potentially belong to the same ungapped alignment.

BLAST-P  detects these  pairs of hits  thanks to the  index previously
construct. All the overlapping words of the sequence from the bank are
considered and search in the look-up table. Interesting  pairs of hits
are  selected if  their distance  is smaller  than a window size of 40
amino acids.

\subsubsection{Stage 2: Ungapped extension}

For each pair  of hits selected in the previous  stage, left and right
ungapped extensions  are started.  This process ignores  insertion and
deletion events. The aim is to  quickly compute a score according to a
given substitution matrix.  Starting from 0, the score is increased or
decreased depending  on the substitution cost between successive amino
acids. The  process start  on the right  hit and  run to the  left (to
reach the  left hit).  It stops  when the score  becomes lower  than a
predefined threshold value or when  it decreases too much. If the left
hit has been reached, then a right extension is run.

If the score exceeds a threshold value $S_{1}$, -- a constant determined
by an external parameter -- then the next stage can be started.

\begin{figure}[h]
  \centering
  \includegraphics[width=3.in]{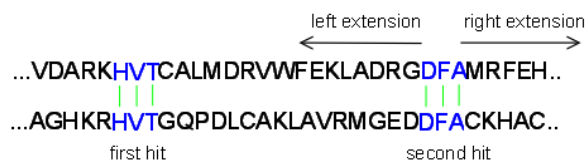}
  \caption{\small \emph{Example about two hits on the same diagonal.}}\label{Fig:TwoHitsDiagonal}
\end{figure}

BLAST-P can  also be run in one  hit mode, where a  single hit, rather
than  two hits,  is required  to trigger  an ungapped  extension. This
leads to an  increase in the number of  ungapped extensions performed,
increasing  runtimes, but  improving  search accuracy.  To reduce  the
number  of  hits,  a larger  value  of  the  neighbor threshold  T  is
typically used when BLAST-P is  run in this mode. The original BLAST-P
algorithm \cite{Altschul90}  was using the  one hit mode.   The double
hit optimization  was one  of the main  changes introduced in  the 1997
BLAST-P \cite{Altschul97}.

\subsubsection{Stage 3: Gapped extension}

In the  third stage, the dynamic  programming algorithm is  used in an
attempt  to  build  gapped  alignments that  passes  through  ungapped
region. A start point is  chosen from ungapped alignment, then dynamic
programming is used to  find the highest-scoring gapped alignment that
passes  through this  point. The  gapped alignment  algorithm  used by
BLAST-P    differs     from    Smith-Waterman    \cite{SMITH}    local
alignment.  Rather  than  exhaustively  computing all  possible  paths
between the sequences, the  gapped scheme explores only insertions and
deletions  that  augment the  high-scoring  ungapped alignment.  After
this, a gapped  alignment is attempted.  The term  gapped alignment is
referred to the approach used  by BLAST-P and local alignment to refer
to the exhaustive Smith-Waterman approach.

A  gapped  alignment  stops  when  the score  falls  below  a  drop-off
parameter,  X.   This parameter  controls  the  sensitivity and  speed
tradeoff: higher the value of X, greater the alignment sensitivity but
slower the  search process.  If  the resulting gapped  alignment score
exceed $S_{2}$ -- which is  determined from an external E-value cutoff
parameter -- it is passed to the final stage.

\subsubsection{Stage 4: Trace-back \& display.}

In   this   stage,   the    final   alignments   are  constructed  and
displayed. During the scoring process, the alignment trace-back pathway
is recorded so  that it can be easily  reconstructed.  Only alignments
that  are considered  as statistically  significant are  output.  More
precisely, only alignments having a score reflecting an expected value
greater than a threshold value set by the user are reported.

\subsection{Code profiling}

In \cite{Cameron04} a   profiling of  the   NCBI BLAST-P  code has been
conducted. It is summarized in the following table.

\begin{table}[h]
\centering\small
\begin{tabular}{|l|c|}
\hline  Task                                                & Percentage of overall time\\
\hline  Find high-scoring short hits                        & 37\% \\
\hline  Identify pairs of hits on the same diagonal         & 18\% \\
\hline  Perform ungapped alignment                          & 13\% \\
\hline  Perform gapped alignment                            & 30\% \\
\hline  Trace-Back and display alignments                   &  2\% \\
\hline
\end{tabular}
\caption{\small \emph{Average runtime for each stage of BLAST-P algorithm on
		Genbank non-redundant protein database \cite{Cameron04}.}}
\label{AverageRuntimeBLASTP}
\end{table}

In this experiment,  the database is the release  113 (August 1999) of
GenBank non redundant protein bank  and 100 queries have been randomly
selected from this bank. The  processor is an Intel Pentium 4 cadenced
at 2.8 GHz  with 1~GByts of main-memory.  BLAST-P has been  run on the
version 2.2.8 with default parameters.

This table shows the percentage of execution time spent on each stage.
It can  be seen  that nearly half  of the  time is spent  in detecting
positions where alignments can be started (stage 1). Our approach aims
at reducing the  time spent during this stage even  if the next stages
will have  to performed more  computation. The idea is  that computing
gapped or ungapped alignments  is much more suited for parallelization
and that, globally, we should gain on the total execution time.

\newpage
\section{PLAST-P algorithm}

\subsection{Introduction}

An immediate way  to speed-up database search is  to split the genomic
database  into  N  parts on  a  N-node  cluster  and to  process  them
separately   as  it   is   done  with   the  mpiBLAST   implementation
\cite{MPI_BLAST}.  A  query sequence is  broadcasted and independently
processed on  each PE  before merging the  results. The  advantages of
this  parallelization are  manifold: access  to the  data is  fast and
efficiency is nearly maximal  since the communication overhead between
the PEs minimal.

For the last 2-3 years,  because of the difficulty of increasing clock
frequency, processor  performance growth has  been limited. To  keep a
high computational power, manufactures now propose chips having several
processor cores.  These new architectures will be  efficient for high
performance computation only if codes fit these architectures.

During  the last decade,  GPUs (graphics  Processing Units)  have been
highly   improved  by   including  a   large  number   of  specialized
processors. The  GPUs have many advantages over  CPU architectures for
processing  highly  parallel  algorithms.   They  now  become  a  real
alternative  for   deporting  very  time   consuming  general  purpose
computations.

To benefit from these last  technologies, we propose a double indexing
bank  approach  allowing  both a  strong  reuse  of  data and  a  high
potential parallelism  on the  new generation of  multicore processors
together with the use of accelerators such as GPU and FPGA.

\subsection{PLAST generic algorithm}

To avoid  the problem of scanning  database many times,  the two banks
are first indexed in the main computer memory before any processing.  Through an
appropriate  index structure,  we  can point  directly  to all  the
identical words (seeds) in both sides of the two banks. If a seed appears
$ |nb1| $ times in bank1 and  $ |nb2| $ times in bank2, then there are
$ |nb1|  \times |nb2|  $ hits. It  means that  there are also  $ |nb1|
\times  |nb2|  $  ungapped   alignments  to  calculate.  The  PLAST-P
algorithm  proceeds in  5 successive  stages.  Stage 0  index the  two
banks, stage  1 constructs two  neighborhood blocks, stage  2 performs
ungapped  extension, stage  3 computes  gapped alignment  and  stage 4
displays  alignments. These  stages  are described in  the
following sections.

\begin{center}
\centering \small{
\begin{tabular}{rll}
\hline \multicolumn{3}{c}{PLAST-P generic algorithm}\\
\hline
 &               index1 = make index (bank1)                                        & \#stage 0 \\
 &               index2 = make index (bank2)                                        &           \\
 &               for all  possible seeds                                            &           \\
 &\hspace{0.40cm}   construct neighborhood  block nb1 from index1                   & \#stage 1 \\
 &\hspace{0.40cm}   construct neighborhood  block nb2 from index2                   &           \\
 &\hspace{0.40cm}   for each subsequence of nb1                                     &           \\
 &\hspace{0.80cm}       for each subsequence of nb2                                 &           \\
 &\hspace{1.20cm}          compute ungapped alignment                               & \#stage 2 \\
 &\hspace{1.20cm}          if score $\geq S_{1}$                                    &           \\
 &\hspace{1.60cm}              compute \textbf{small} gapped alignment              &\#stage 3.1\\
 &\hspace{1.60cm}              if score $\geq S_{2}$                                &           \\
 &\hspace{2.00cm}                  compute \textbf{full} gapped alignment           &\#stage 3.2\\
 &\hspace{2.00cm}                  if score $\geq S_{3}$                            &           \\
 &\hspace{2.40cm}                      trace-back \& display alignment              & \#stage 4 \\
\hline
\end{tabular}}
\end{center}

\subsection{The 5 stages of the PLAST-P algorithm}

\subsubsection{Stage 0: Bank indexing.}

In the PLAST-P algorithm, bank indexing is done using the subset seed
concept   \cite{Kucherov06}  \cite{Noe05}.    Subset  seeds   have  an
intermediate  expressiveness  between  spaced  seeds  \cite{Li04}  and
vector  seeds  \cite{Daniel05}.  Their  main  advantage  is that  they
provide a powerful seed definition while preserving the possibility of
direct indexing. Here, we use the following subset seed:

\begin{table}[h]
\small
\begin{tabular}{l}
A,C,D,E,F,G,H,K,L,M,N,P,Q,R,S,T,V,W,Y            \\
c=\{C,F,Y,W,M,L,I,V\}, g=\{G,P,A,T,S,N,H,Q,E,D,R,K\} \\
A,C,f=\{F,Y,E\},G,i=\{I,V\},m=\{M,L\},n=\{N,H\},P,q=\{Q,E,D\},r=\{R,K\},t=\{T,S\} \\
A,C,D,E,F,G,H,K,L,M,N,P,Q,R,S,T,V,W,Y            \\
\end{tabular}
\end{table}

\begin{figure}[h]
  \centering
  \includegraphics[width=4.in]{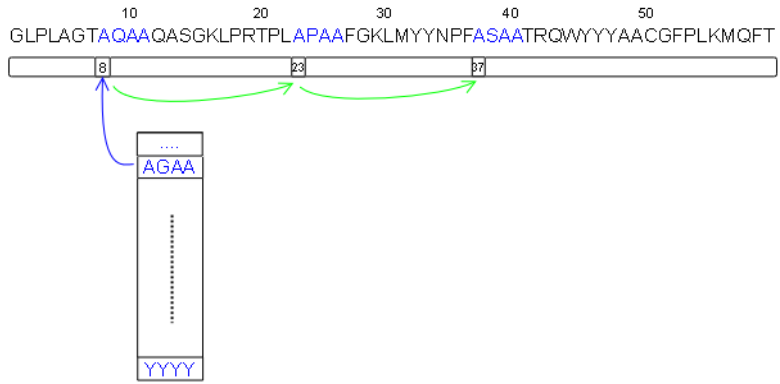}
  \caption{\small \emph{Index structure for storing the positions of the word AGAA.}}
  \label{Fig:IndexStructure}
\end{figure}

The index structure of a bank is shown figure \ref{Fig:IndexStructure}.
All words of length W  (seeds) are considered and translated into
their  corresponding  subset seed. For  example, the words  AQAA, APAA,
ASAA     are     translated     to     a     unique     word:     AgAA.
Then,  words having the same  subset seed are linked together.

A look-up table containing all the subset seeds gives a direct access
to the  linked lists. For each  subset seed, its first  occurrence in
the database is provided.

\subsubsection{Stage 1: Neighborhood block building}

Hits detected in the first  stage are the starting point for computing
ungapped alignments.  The objective is to be able to rapidly decide if
one hit has  favorable environment to build an  alignment. The index
structure gives the position of  the W-AA words, allowing to access to
the neighboring amino acids for processing ungapped alignments.

\begin{figure}[h]
  \centering
  \includegraphics[width=4in]{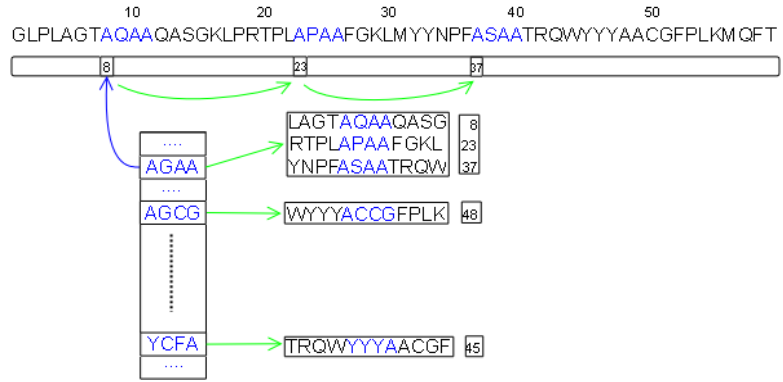}
  \caption{\small \emph{Neighborhood blocks including seeds and 4 characters on their left and right hand side }}\label{Fig:ConstructSubsequence}
\end{figure}

In this stage,  for all identical seeds,  two neighborhood blocks
are  built: one  block (nb1)  for bank1  and another  block  (nb2) for
bank2. A  block is made of  neighborhood amino acids  around each seed.
Building blocks allow  a best  reuse of  the cache  memory when
computing  ungapped alignment.   Figure \ref{Fig:ConstructSubsequence}
shows the  blocks of subsequences for  words AGAA, AGCG  and YCFA with
four neighborhood amino acids on the left and on the right.

\subsubsection{Stage 2: Ungapped alignment.}

For  each different  seed, we  perform $|nb1|  \times  |nb2|$ ungapped
alignments  from the  two neighborhood  blocks previously  built. More
precisely,  for  each  subsequence  in  block nb1,  we  calculate  the
ungapped  score  with  $|nb2|$  subsequences  in  block  nb2.   Unlike
BLAST-P, ungapped  extension is  performed on subsequences  of bounded
length. Another major difference  is that ungapped extension starts as
soon as  a single hit is identified (BLAST-P requires two  hits on the
same  diagonal   and  closed  to  each  another   before  starting  an
alignment). If the  score  of the ungapped alignment  is greater than a
threshold value $S_{1}$, it is passed to the next stage.

According to our experimentations,  the computation time of this stage
is very  important. This  part of code is thus a strong  candidate for
parallelization.

The  ungapped  algorithm  is the  following: starting  from the  first
character  of seed  on both subsequences, the  extension  runs on  the
right. It  stops  when  the  score  becomes  lower  than a  predefined
threshold  value or  when it decreases too much. If the score dose not
exceeds a threshold value $S_{1}$ then a left extension is run.

\subsubsection{Stage 3: Gapped alignment.}

Dynamic programming  technique is used to  extend ungapped alignments.
Actually, this  stage is divided  into two sub-stages: the  first find
small gapped alignments  (stage 3.1).  It aims to  limit the searching
space  of dynamic  programming by  allowing only  a maximal  number of
gaps.  If  the score exceeds  a threshold value $S_{2}$,  the standard
dynamic  programming procedure  is launched  (stage  3-2). Experiments
have shown that, in many cases, the second step is not done (90\%).

The reason why this stage is  divided into two parts is that the first
one is  much more suited  for parallelization on  hardware accelerator
and  appears  as  a  good   filter  before  starting  a  full  dynamic
programming search.

Most of  the time,  gapped alignments content  more than  one ungapped
alignment. In such a case, two ungapped alignments are not on the same
diagonal.   For  this reason,  sometime,  same  gapped alignments  are
recalculated  from  different  ungapped  alignments.   To  avoid  this
problem, all  ungapped alignments  belonging to gapped  alignments are
stored in a structured list.  Before starting the computation of a new
gapped alignment, we first check if the hit doesn't belong to ungapped
alignment of this list.

\begin{figure}[h]
  \centering
  \includegraphics[width=3in]{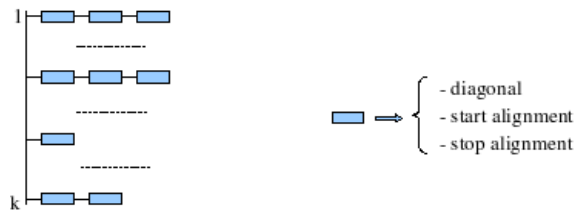}
  \caption{\small \emph{Linked lists of three ungapped alignment fields.}}
  \label{Fig:ListsUngappedAlignments}
\end{figure}

In the list, for each ungapped alignment, three fields are stored: the
diagonal number, the start position and the stop position. Actually, a
linked   lists  is   used  to   store  alignments   as   shown  Figure
\ref{Fig:ListsUngappedAlignments}.   There are  k  linked lists,  each
linked list storing a group of diagonal.

\subsubsection{stage 4: Trace-back \& display.}

In  this  stage,   trace-back  information  optimizes  the  alignments
recorded in  the previous stage and  displays them to the  user. It is
similar to The NCBI BLAST-P processing.

\newpage
\section{PLAST-P sequential version}

\subsection{Code profiling}

The   execution   time  of   each   stage   are   detailed  in   Table
\ref{AverageRuntimePLASTP}. In this experiment, data input are {\sc bank1}
and  {\sc bank2}  (cf. Annex 1).  Stage 2  and  stage 3.1 respectively
consumed an average of 71.8\% and 17.5\% of the total execution time.

\begin{table}[h]
\centering\small
\begin{tabular}{|c|l|c|}
\hline Stage & Task                                        & Percentage of overall time\\
\hline 0     & Index the two banks                                & 0.1\%   \\
\hline 1     & Construct neighborhood blocks               & 0.2\%  \\
\hline 2     & Perform ungapped alignment                  & 71.8\%  \\
\hline 3-1   & Perform small gapped alignment              & 17.5\%  \\
\hline 3-2   & Perform full gapped alignment               & 7.9\%   \\
\hline 4     & Trace-Back information and display alignments& 2.4\%   \\
\hline
\end{tabular}
\caption{\small \emph{Average runtime for each stage of PLAST-P algorithm when taking as input {\sc bank1} and \sc{bank2}.}}
\label{AverageRuntimePLASTP}
\end{table}

The  processing times  of both  ungapped and  small  gapped alignments
represent the  majority of  the runtime.  Thus,  to get  a significant
speed up, both stages need to be optimized. To improve the performances
of these stages, two methods have been implemented. The first  one  is
based on the  distribution  of values in the substitution  matrix, and
acts as a filter which eliminates hits generating potentially not good
alignments. The second approach uses Single-Instruction  Multiple Data
(SIMD), a technique  employed to achieve data level parallelism.

\subsection{Filter optimization}

The filter aims to eliminate  hits providing not a good environment to
produce significant ungapped alignments.  Its function is based on the
distribution of scores which are  greater than zero in the substitution
matrix (about  15\%). A score  between two sequences is  calculated as
the sum of all amino acids pairwise having a matrix score greater than
zero.  Thus, $|nb1|  \times |nb2|$ scores are computed  and only pairs
of sequences  exceeding a threshold  values $S_{0}$ are  considered as
good candidates for the next stage.

To  implement  this  filter,  a  three-dimensional  table  stores  the
positions of scores greater than zero between all amino acids in block
nb1. The three dimensions correspond respectively to (1) the number of
elements in block nb1, (2) the  length of the subsequences and (3) the
number of amino-acid (20).

\begin{figure}[h]
  \centering
  \includegraphics[width=4in]{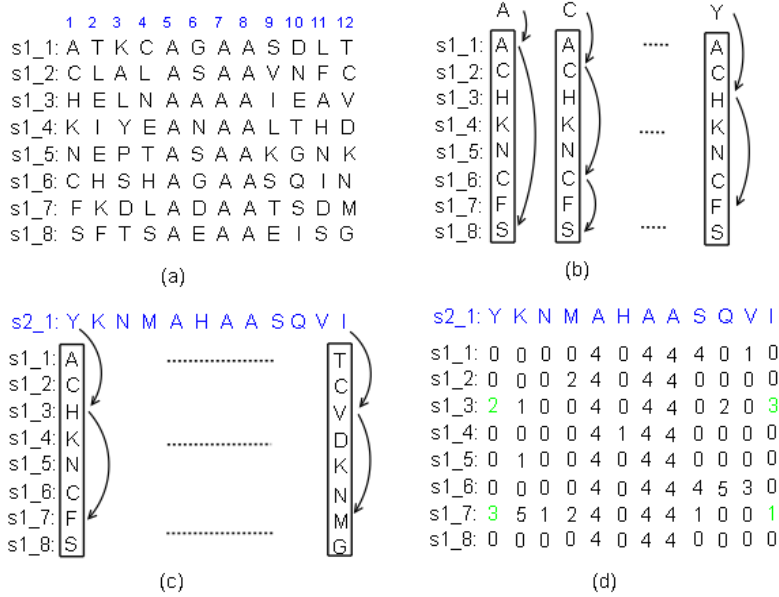}
  \caption{\small \emph{Filtering based on the calculation of a raw score for subsequences of 12 amino-acids.}}
  \label{Fig:SumScoreSuperior}
\end{figure}

One example is shown Figure \ref{Fig:SumScoreSuperior}: nb1 contains 8
subsequences  (s1\_1,  s1\_2, ...,  s1\_8)  of  length  12.  For  each
character position  in the subsequences,  20 linked lists  are created
(they correspond   to the 20  amino acids). Each linked  list contents
the positions of  the amino acid where the score  is superior to zero.
The next step  is to compute the score  between subsequence (s2\_1) in
block nb2  with   the  8 subsequences  in  nb1. To do this,  for  each
amino acid in subsequence s2\_1, one linked list corresponding to this
acid amino  is chosen (Figure  \ref{Fig:SumScoreSuperior}.c). Based on
the elements  in this  linked list, we  calculate a score  as depicted
Figure \ref{Fig:SumScoreSuperior}.d.

To evaluate the filter, the  protein {\sc bank1}  and {\sc bank2} (cf.
Annex 1) have been considered  as input data.  Measures  indicate that
the filter  can reduce the execution  time of 50\%  compared to direct
ungapped alignment. Only 2,5\% of the hits are really processed by the
original  ungapped  procedure.  In  that  scheme,  the filtering  time
represents about 86\% of the overall ungapped alignment stage.

Table \ref{AverageRuntimePLASTPFilter}  shows the new  code  profiling
when the filter is activated. It can be seen that the second and third
(3-1) stages  are still  time  consuming (82\% of  the total execution
time).

\begin{table}[h]
\centering\small
\begin{tabular}{|c|l|c|}
\hline Stage & Task                                        & Percentage of overall time\\
\hline 0     & Index the two banks                              & 0.1\%  \\
\hline 1     & Construct neighborhood blocks               & 0.2\%  \\
\hline 2     & Perform ungapped alignment                  & 55.8\% \\
\hline 3-1   & Perform small gapped alignment              & 26.4\% \\
\hline 3-2   & Perform full gapped alignment               & 11.8\% \\
\hline 4     & Trace-Back and display alignments            & 3.6\%  \\
\hline
\end{tabular}
\caption{\small \emph{Average runtime for each stage of PLAST-P algorithm taking as input {\sc bank1} and {\sc bank2} with filter optimization.}}
\label{AverageRuntimePLASTPFilter}
\end{table}

\subsection{SIMD optimization}

\subsubsection{Ungapped alignments}

Generally, the scores of ungapped  alignments are small since they are
computed on  subsequences of  limited size. Hence,  they need  a small
number of  bits to store the  integer values. Here, 1  byte is  enough
for allowing a 128-bit SIMD register to contain 16 scores.

\begin{figure}[h]
\small
\begin{verbatim}
                 block nb1                                block nb2
    s1_01: .. K C A G A A S D ..             s2_01: .. N M A H A A S Q ..
    s1_02: .. A G A S A A V N ..
    s1_03: .. L N A A A A W E ..

    ...

    s1_14: .. S H A G A A S Q ..
    s1_15: .. W Q A D A A T S ..
    s1_16: .. T S A E A A E M ..
\end{verbatim}
\normalsize
\end{figure}

In our implementation, the idea is to compute in parallel the score of
16  subsequences from block  nb1 with  1 subsequence  of block  nb2 as
shown on the above figure. If  the score exceeds the 8-bit range, it is
tagged as an overflow value and passed to next stage.

\subsubsection{Gapped alignments}

SIMD instructions are  also used for speeding up  the execution of the
small gapped  alignment step.   To exploit SIMD  parallelism, ungapped
alignments  (coming  from  the previous  step) are  first stored  in a
list.  When the  list contains at least K  elements, they are processed
in a SIMD fashion for computing gapped extension.

Unlike ungapped alignments, the  score of  gapped extensions generally
exceed 255 (8-bit). Thus, the SIMD registers use a 16-bit partitioning.

\small
\begin{verbatim}
                   bank1                                 bank2
   s1_1: P L..K S A G A A S G..S R       s2_1: R N..A D A H A A G P..K R
   s1_2: I K..R N A S A A F K..Y L       s2_2: F E..Q Q A G A A F K..Y Q
   s1_3: F I..V R A A A A K Q..L I       s2_3: F F..G Q A N A A K I..Q A
   s1_4: D F..I V A N A A Q D..P K       s2_4: D F..Y G A P A A I F..A K
   s1_5: G D..R I A S G A E L..S D       s2_5: L Q..L I A K A A E V..M E
   s1_6: Q G..H R A G A A L S..D I       s2_6: Q L..H L A R A A V N..E K
   s1_7: K S..D L A D A A E I..F S       s2_7: L T..N F A N A A D F..F S
   s1_8: S F..R S A E A A Q N..I L       s2_8: G P..G F A T A A Q N..L L
\end{verbatim}
\normalsize

Here, the SIMD implementation considers the computations of 8 alignments
in parallel, each  of  one  performing  the  Smith-Waterman  algorithm
restricted to a limited diagonal \cite{SMITH}. The figure above illustrates
the data involved in one iteration. The SIMD algorithm can be found in
Annex 3.

To evaluate the benefit of the SIMD parallelism, the protein {\sc bank1}
and {\sc bank2} (cf. Annex 1) have been considered as input  data. For
ungapped (respectively small gapped)  alignments, SIMD  implementation
achieves  a  speed-up  ranging  from 4  to 6 (respectively  2.5  to 3)
compared to the filter implementation.

Table \ref{AverageRuntimePLASTPSIMD} shows  that the second and third
(3-1) stages still consumed about 57.5\% of the total execution time.

\begin{table}[h]
\centering\small
\begin{tabular}{|c|l|c|}
\hline Stage & Task                                        & Percentage of overall time\\
\hline 0     & Index the two banks                     & 0.4\%  \\
\hline 1     & Construct neighborhood blocks               & 0.7\%  \\
\hline 2     & Perform ungapped alignment                  & 33\%   \\
\hline 3-1   & Perform small gapped alignment              & 24.5\% \\
\hline 3-2   & Perform full gapped alignment               & 32.1\% \\
\hline 4     & Trace-Back and display alignments            & 9.3\%  \\
\hline
\end{tabular}
\caption{\small \emph{Average runtime for each stage of PLAST-P algorithm taking as input {\sc bank1} and {\sc bank2} with SIMD 16-bit optimization.}}
\label{AverageRuntimePLASTPSIMD}
\end{table}

\subsection{Perfornamces}

In this section, we report measures of the different  implementations
of PLAST-P. For each of them, the same data set is  used: {\sc bank1}
(141.7K sequences) and {\sc bank2} (5K, 10K, 20K, 40K sequences).

\subsubsection{Total execution time}

Table \ref{PerformancePLASTPBLAST-P} reports the total execution time
of NCBI BLAST-P and four  implementations of PLAST-P. The last column
is dedicated to  the NCBI BLAST-P using default  parameters (e-value =
0.001).  The  second column  (PLAST-P -- no  opt) corresponds  to the
program PLAST-P with no  optimization.  The third column (PLAST-P --
filter)  includes the filter  optimization. The  two next  columns are
related  to  the  SIMD  implementation,  first,  with  SIMD  registers
partitioned into  8 16-bit  registers  (PLAST-P -- SIMD 16  bit) and,
second, with  SIMD registers  partitioned into 16  8-bit registers for
the ungapped extension stage only.

\begin{table}[h]
\centering
\small
\begin{tabular}{|c|c|c|c|c|c|}
\hline nb. seq.    & PLAST-P     &  PLAST-P   &   PLAST-P     & PLAST-P     & BLAST-P \\
       bank2       & (no opt.)    & (filter)    & (SIMD 16-bit)  & (SIMD 8-bit) & (NCBI)  \\
\hline 5k          &  3,766	      &  2,570	    &      940  	 &     857      &  3,521  \\
\hline 10k         &  7,188	      &  4,741  	&    1,775	     &   1,585      &  6,832  \\
\hline 20k         & 14,394       &  9,333      &    3,489       &   3,188      & 13,597  \\
\hline 40k         & 28,397       & 18,996      &    6,677       &   6,053      & 26,111  \\
\hline
\end{tabular}
\caption{\small \emph{Execution time (sec) of BLAST-P and various versions of PLAST-P.}}
\label{PerformancePLASTPBLAST-P}
\end{table}

\begin{table}[h]
\centering
\small
\begin{tabular}{|c|c|c|c|c|}
\hline nb. seq.    & PLAST-P     &  PLAST-P     &   PLAST-P      & PLAST-P       \\
       bank2       & (no opt.)   & (filter)     & (SIMD 16-bit)  & (SIMD 8-bit)  \\
\hline 5k          &   0.93      &  1.37       &    3.74         &    4.1         \\
\hline 10k         &   0.95      &  1.44       &    3.84         &    4.3         \\
\hline 20k         &   0.94      &  1.45       &    3.89         &    4.26         \\
\hline 40k         &   0.91      &  1.37       &    3,91         &    4.31         \\
\hline
\end{tabular}
\caption{\small \emph{Speed-up of PLAST-P compared to BLAST-P.}}
\label{SpeedupPLASTPBLAST-P}
\end{table}

From Tables \ref{PerformancePLASTPBLAST-P} and \ref{SpeedupPLASTPBLAST-P},
it   can   be   seen   that   the   SIMD  implementation is much  more
efficient than the filter optimization. It achieves a global  speed up
of 4.3 compared to NCBI BLAST-P.

\subsubsection{Execution time of ungapped alignments}

As the ungapped extension step is very time consuming, it is important
to have a closer look on the way the different implementations improve
this critical part. Table \ref{ComparasionUGPLAST-P}
reports  the different  execution times. The 8-bit SIMD implementation
provides a speed-up of 12 compared to the original code.

\begin{table}[h]
\centering
\small
\begin{tabular}{|c|c|c|c|c|}
\hline nb. seq.        &  PLAST-P & PLAST-P       & PLAST-P       & PLAST-P         \\
	bank2              & (no opt.) & (filter)       & (SIMD 16-bit)  & (SIMD 8-bit) \\
\hline 5k              &  2,741    & 1,511          &     312        &    230       \\
\hline 10k             &  5,195    & 2,643          &     600        &    423       \\
\hline 20k             & 10,187    & 5,102          &   1,204        &    830       \\
\hline 40k             & 20,259    &10,256          &   2,237        &   1,534      \\
\hline
\end{tabular}
\caption{\small \emph{Execution time (sec) of ungapped alignment in different implementations.}}
\label{ComparasionUGPLAST-P}
\end{table}

\begin{table}[h]
\centering
\small
\begin{tabular}{|c|c|c|c|}
\hline nb. seq.        & PLAST-P        & PLAST-P        & PLAST-P      \\
	bank2              & (filter)       & (SIMD 16-bit)  & (SIMD 8-bit) \\
\hline 5k              & 1.81           &     8.78       &   11.91      \\
\hline 10k             & 1.96           &     8.65       &   12.28      \\
\hline 20k             & 1.99           &     8.46       &   12.27      \\
\hline 40k             & 1.95           &     9.05       &   13.02      \\
\hline
\end{tabular}
\caption{\small \emph{Speed-up of ungapped alignment compared to no optimization.}}
\label{SpeedupUGPLAST-P}
\end{table}

\subsubsection{Selectivity of stage 2}

The stage 2 acts  as a filter to eliminate hits which  have not a good
chance to  provide significant alignments.  Hits are discarded  if the
score of ungapped alignments are  lower than a threshold value. We can
then measure the percentage of hits which pass successfully this stage,
as it is shown table \ref{ComparasionNbPLAST}. Lower this percentage,
smaller  the   number  of  gap   extensions  to  perform.   From  table
\ref{ComparasionNbPLAST}, it can be seen that the filter optimization
loses a few hits compared to other implementations.

\begin{table}[h]
\centering
\small
\begin{tabular}{|c|c|c|c||c|c|c|}
\hline   nb. seq.  & \multicolumn{3}{|c||}{\% successful ungapped extension}& \multicolumn{3}{|c|}{number of alignments found}\\
\cline{2-7}bank2       & (no opt.)  & (filter)    & (SIMD)     & (no opt.)  & (filter)    &(SIMD)  \\
\hline 5k              & 0.186\%   & 0.166\%    & 0.186\%   & 305,493   &  305,044    &   305,498  \\
\hline 10k             & 0.185\%   & 0.165\%    & 0.185\%   & 611,005   &  610,543    &   611,093  \\
\hline 20k             & 0.185\%   & 0.165\%    & 0.185\%   &1,059,776  &1,058,660    & 1,059,822  \\
\hline 40k             & 0.184\%   & 0.165\%    & 0.184\%   &2,237,061  &2,235,328    & 2,237,122  \\
\hline
\end{tabular}
\caption{\small \emph{Results of different PLAST-P sequential implementations.}}
\label{ComparasionNbPLAST}
\end{table}

\subsubsection{Sensitivity}

Based on the definition of  two equivalent alignments  (see cf.  Annex
2), we  compared   the  sensitivity   between  PLAST-P  and    BLAST-P
implementations. Here, the 8-bit SIMD optimization  is only considered
as it provides the best  performances. The sensitivity is evaluated by
considering three values of \%~margin: 2\% 5\%  and 10\% (cf Annex 2).
Table \ref{SensitivityPLASTPBLAST-P} summarizes the results.  For each
test, the number of  alignments found  by  BLAST-P and SIMD PLAST-P is
specified as a percentage of equivalent alignments.

\begin{table}[h]
\centering
\small
\begin{tabular}{|c|c|c|c|c|c|}
\hline   nb. seq & nb. alig. BLAST-P & nb. alig. PLAST-P& 2\%    &    5\% &   10\%         \\
\hline 5k                 &   305,435        &   305,489         & 94.7\% & 94.9\% & 95.4\% \\
\hline 10k                &   611,031        &   611,093         & 95.3\% & 95.3\% & 95.8\% \\
\hline 20k                & 1,047,794        & 1,059,822         & 94.8\% & 95.0\% & 95.4\% \\
\hline 40k                & 2,237,076        & 2,237,122         & 94.5\% & 95.2\% & 95.6\% \\
\hline
\end{tabular}
\caption{\small \emph{Comparison of sensitivity between PLAST-P and BLAST-P}}
\label{SensitivityPLASTPBLAST-P}
\end{table}

Both programs (BLAST-P  and PLAST-P) detect  about the  same number of
alignments, and approximately 95  \% of the alignments are equivalent.
The difference  can be explained  as both programs don't  consider the
same  seeds to detect  hits. Thus,  there are  some  alignments  found
by BLAST-P and not by PLAST-P and, on  the  contrary,  some alignments
found  by  PLAST-P  and    not  by  BLAST-P.  As  an  example,  Figure
\ref{Fig:AlignmentsFound}  presents  two  alignments  which  have  not
been detected by each program. The first alignment is found by PLAST-P
and  not by  BLAST-P because it doesn't exist  a  pair  of hits on the
same  diagonal. The  second  alignment is found by BLAST-P, but not by
PLAST-P because the subset  seed system of PLAST-P did not detect such
hits.

\begin{figure}[h]
{\it Alignment 1 (found by PLAST-P)}
\footnotesize
\begin{verbatim}
     I980: 112  STKIMKSAIIADSATIGKNCYIGHNVVIEDDVIIGDNSIID----AGTFIGRGVNIGKNA  167
                + KI  SA+I  + TIG N  +G    I+  V++ ++ + D      T +G    IGK A
     P190: 261  TAKIHPSALIGPNVTIGPNVVVGEGARIQRSVLLANSQVKDHAWVKSTIVGWNSRIGKWA  320
     I980: 168  RIEQHVSINYAIIGDD--V---VILVGAK  191
                R E        ++GDD  V   + + GAK
     P190: 321  RTE-----GVTVLGDDVEVKNEIYVNGAK  344
\end{verbatim}
\normalsize
{\it Alignment 2 (found by BLAST-P)}
\footnotesize
\begin{verbatim}
     I914: 109  HSCFNMSSSVMKQMRNQNYGRIVNISSINAQAGQIGQTNYSAAKAGIIGFTKALARETAS  168
                     M  + +  M+ +  GR++   S+    G      Y A+K  + G  ++LA
     1FDV: 116  VGTVRMLQAFLPDMKRRGSGRVLVTGSVGGLMGLPFNDVYCASKFALEGLCESLAVLLLP  175
     I914: 169  KNITVNCIAPGYIATEMVNTV---PKDILTK  196
                  + ++ I  G + T  +  V   P+++L +
     1FDV: 176  FGVHLSLIECGPVHTAFMEKVLGSPEEVLDR  206
\end{verbatim}
\caption{\small \emph{Example of two alignments: (1) found by PLAST-P; (2) found by BLAST-P.}}
\label{Fig:AlignmentsFound}
\normalsize

\end{figure}

\newpage
\section{PLAST-P parallel version}

This section  presents three parallel version of  PLAST-P.  The first
one -- multicore PLAST-P -- works on multicore processors, the second
one -- GPU  PLAST-P -- uses graphic processors  to accelerate stage 2
and  stage 3-2  and the  final version  -- multicore  GPU  PLAST-P --
combines the two approaches.

\subsection{Multicore processors}

On the PLAST-P algorithm, it can be noticed that stages 1, 2 and 3 can
be computed independently  for each seed. Thus, if M is the  number of
all the possible seeds, then  the  PLAST-P  algorithm can be  splitted
into M different processes working separately.

As the number of available CPUs on a  multicore processor is much less,
one CPU will have -- on average -- the charge of computing $\frac{M}{P}$
tasks. For  load  balancing  purpose, a  CPU is not initialized with a
predefined list of tasks. It is just initialized with one task assigned
to  one  specific seed. When its task is finished, it asks for another
seed to process. The program stops when there are no more tasks to
process.

In summary, the sequential algorithm is changed into a parallel algorithm
as follows:

\begin{center}
\small
\begin{tabular}{ll}
\hline\hline \multicolumn{2}{c}{PLAST-P parallel algorithm on multicore processors}   \\
\hline \multicolumn{2}{c}{\emph{Main thread}}                                         \\
\hline
                index1 = make index (bank1)                           &   \#stage 0   \\
                index2 = make index (bank2)                           &               \\
                initialize the shared index variable idx = 1          &               \\
                create P slave threads                                &               \\
                wait slave threads to finish and join results         &               \\
\hline
\multicolumn{2}{c}{\emph{Slave threads}}                                              \\
\hline
               while ( idx $\leq$ M ) do                             &               \\
\hspace{0.40cm}    pthread\_mutex\_lock()                            &               \\
\hspace{0.80cm}        get a seed from idx                           &               \\
\hspace{0.80cm} 	    modify idx (idx++)                           &               \\
\hspace{0.40cm} 	pthread\_mutex\_unlock()                         &               \\
\hspace{0.40cm} 	construct two neighbouring blocks (nb1 \& nb2)   &   \#stage 2   \\
\hspace{0.40cm}    for all hits between nb1 and nb2                  &               \\
\hspace{0.80cm}        compute ungapped alignment                    &               \\
\hspace{0.80cm}        if score $\geq S_{1}$                         &               \\
\hspace{1.20cm}            compute \textbf{small} gapped alignment   &   \#stage 3.1 \\
\hspace{1.20cm}            if score $\geq S_{2}$                     &               \\
\hspace{1.60cm}                compute \textbf{full} gapped alignment&   \#stage 3.2 \\
\hspace{1.60cm}                if score $\geq S_{3}$                 &               \\
\hspace{2.00cm} 	                store them in T\_ALIGN[$P_{i}$]  &               \\
               end while                                             &               \\
               trace-back alignments in T\_ALIGN[$P_{i}$]            &   \#stage 4   \\
\hline\hline\\
\end{tabular}
\end{center}

The implementation  has been done  with the pthread  library: the main
thread  performs  sequentially stage 0, then  creates P slave  threads
to perform in parallel  stages 1, 2 and 3 for  each  seeds. Alignments
detected by thread $P_{i}$ are stored in its private T\_ALIGN structure.
When there are no more seed task to process, each threads perform stage
4 for alignments in their T\_ALIGN structure.

\subsection{GPU - Graphics Processing Units}

During the last decade, GPUs \cite{userguideCUDA} have  been developed
as highly  specialized  processors for the  acceleration of  graphical
processing. The GPUs have  several advantages  over CPU  architectures
for highly  parallel  intensive  workloads,  including  higher  memory
bandwidth,  significantly  higher  floating-point   capability,    and
thousands of  hardware  thread  contexts  with  hundreds  of  parallel
processing units executing  programs in a single  instruction multiple
data (SIMD) fashion.

Recently (2007),  NVIDIA has  introduced  the Geforce  8800 GTX  board
together with a C-language programming   called  CUDA  \cite{userguideCUDA}
(Compute Unified Device Architecture). Geforce 8800  GTX  architecture
comprises 16 multiprocessors. Each multiprocessor has 8 SPs (Streaming
Processors) for  a total of 128 SPs. Each group of 8 SPs shares one L1
data cache. A  SP contains  a scalar  ALU (Arithmetic  Logic Unit) and
can perform floating point operations. Instructions  are executed in a
SIMD mode.

On the GPU,   threads are organized in blocks. A grid of thread blocks
is  executed  on the device. Thread blocks have the same dimensions, and
are processed  by only  one multiprocessor, so  that the
shared  memory space  resides in the on-chip shared  memory leading to
very  fast  memory  accesses. A  multiprocessor  can  process  several
thread  blocks concurrently by partitioning among the set of registers
and the shared memory.

For  each seed, $|nb1| \times |nb2|$ ungapped alignments are performed
in parallel  by the GPU. The  ungapped extensions passed to the  stage
3.1 are stored in a list. When this list contains at least K elements,
all elements  on this list are  considered for small gapped alignment,
again on the GPU.

\subsubsection{Ungapped alignment}

\begin{figure}[h]
  \centering
  \includegraphics[width=2.5in]{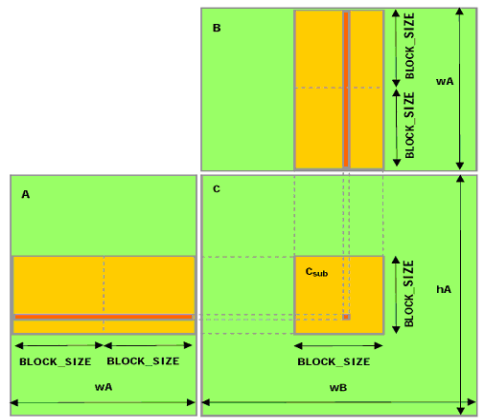}
  \caption{\small \emph{Computing ungapped alignment on GPU.}}
  \label{Fig:UngappedAlignmentGPU}
\end{figure}

We  implemented the ungapped  alignment stage  by adapting  the matrix
multiplication   algorithm    given   in   the    CUDA   documentation
\cite{userguideCUDA}. For  each   seed, there   are  two   subsequence
blocks. Suppose that block A[wA,  hA] corresponds  to block nb1: wA is
the length of  subsequences, and  hA is the  number of subsequences in
block nb1, and block  B[wB, hB]  corresponds  to block nb2:  wB is the
number   of   subsequences in   block nb2,   and hB   is the length of
subsequences. Actually, Block  B  is the  transposition  of block nb2.
Furthermore,  we use  an  other block  C[hA,  wB] to  store scores  of
ungapped alignments between block nb1 and block nb2. The value of each
cell[i,j] in block C corresponds to  the score of subsequence j (row j
of block nb1) and subsequence i (column i of block nb2).

The task  of $|nb1| \times |nb2|$ ungapped  alignments  between  block
A and block B is split among threads on GPU as followed:  each  thread
block  is  responsible for  computing a square sub\-block $C_{sub}$ of
C. Each  thread  within  the  block  is responsible for computing  one
element  of  $C_{sub}$  (Figure  \ref{Fig:UngappedAlignmentGPU}).  The
dimension  block\_size of $C_{sub}$ is chosen to be equal to 16. Thus,
there are  $\frac{hA}{16}$  x  $\frac{wB}{16}$ thread  blocks in grid.
Threads within the same block share the same memory space.

Two subsequence  blocks are  mapped to the  texture memory of the GPU.
The texture memory is shared by all the processors, and speed up comes
from its  space  which is  implemented as a  read-only  region  of the
device memory. At the beginning of the computation, each  thread loads
a character  from the  texture  memory to  the shared  memory  using a
texture reference, called texture fetching.

\subsubsection{Small gapped alignments}

\begin{figure}[h]
  \centering
  \includegraphics[width=2.5in]{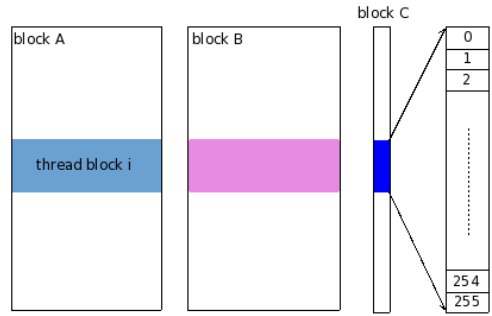}
  \caption{\small \emph{Computing small gapped alignment on GPU.}}
  \label{Fig:smallGapGPU}
\end{figure}

The use  of high  performance  computing  on GPU  is efficient only to
perform  large  tasks. Thus,  we use the GPU  to execute small  gapped
alignments when there are at least K elements ready  for  computation.
With K small gapped  alignments, there are  K extensions  -  extending
in two directions. To be able  to compute 2K extensions on the GPU, we
have to construct two subsequence blocks as already done  for ungapped
alignment: one  block (A)  for  bank1 and  one  block  (B) for  bank2.
Compared to ungapped alignment, there  is a  difference: for one small
gapped alignment, we copy  two subsequences in bank1 - one at the left
and one at the right of the seed - to block  A, and  two  subsequences
(bank2) to  block  B. Consequently, there are 2K  subsequences in each
block. The GPU divides 2K extensions into  thread  blocks; each thread
within a block is responsible for computing one extension.

The form of thread block is $1 \times 256$. Thus, there are $\frac{2K}
{256}$ thread blocks. Two subsequence blocks are mapped to the texture
memory. At the beginning of the  computation, each  thread  copies its
pair of subsequences from the texture memory by the  texture reference
to its local memory for reducing memory access conflict. The scores of
2K extensions are stored in block C[2K,1] as shown Figure \ref{Fig:smallGapGPU}.

\subsection{Multicore processors and GPU}

This section describes a parallelization based on a platform made of a
single   dual  core   processor   and  a   single   GPU  board.    The
parallelization works  as follows: the  first core constructs  the two
neighborhood blocks and controls the computation -- ungapped alignment
and small  gapped alignment  -- on the GPU.  The second  core performs
full gapped alignment and stage  4. The two cores communicate together
by a queue  Q.  The multicore GPU parallel  algorithm can be described
as follows:\\

\begin{center}
\small
\begin{tabular}{rll}
\hline\hline \multicolumn{3}{c}{Multicore GPU PLAST-P algorithm}                            \\
\hline
 &              index1 = make index (bank1)  		                            & \#stage 0  \\
 &              index2 = make index (bank2)                                     &            \\
\hline \multicolumn{3}{c}{Processor 1}                                                       \\
\hline
 &              for all possible seeds                                          &            \\
 &\hspace{0.40cm}    construct neighborhood  block nb1 from index1              & \#stage 1  \\
 &\hspace{0.40cm}    construct neighborhood  block nb2 from index2 		        &            \\
 &\hspace{0.40cm}    compute $|nb1| \times |nb2|$ ungapped alignments on GPU	& \#stage 2  \\
 &\hspace{0.40cm}    store ungapped alignments if score $\geq S_{1}$ in list R  &            \\
 &\hspace{0.40cm}    if the list R contains at least K elements 				&            \\
 &\hspace{0.80cm}       compute K small gapped alignments on GPU 				& \#stage 3.1\\
 &\hspace{0.80cm}       store small gapped alignments if score $\geq S_{2}$ in queue Q     & \\
 &\hspace{0.80cm}       deleted K elements in R 						        &            \\
 &              flag = FALSE									                &            \\
\hline \multicolumn{3}{c}{Processor 2}								                         \\
\hline
 &while(flag==TRUE)										                        &            \\
 &\hspace{0.40cm}   if list Q != $\varnothing$						            &            \\
 &\hspace{0.80cm}       compute full gapped alignment for all elements in Q	    & \#stage 3-2\\
 &\hspace{0.80cm}       if score $\geq S_{3}$						            &            \\
 &\hspace{1.20cm}           trace-back display alignment                        & \#stage 4  \\
\hline\hline
\end{tabular}
\end{center}

\subsection{Perfornamces}

In  this  section, we  report  measures  of  the  different  parallel
implementations of PLAST-P.   For each of them, the  same data set is
used: {\sc bank1} and {\sc bank2}(cf. Annex 1).

The multicore processors  version has been tested with  the SIMD 8-bit
optimization.  The GPU version  uses the  NVIDIA GeForce  8800 graphic
board  (cf. Annex  1).   The multicore-GPU  version  combines the  two
approaches: dual core processor + GPU board.

\subsubsection{Total execution time}

The execution time is compared between the different parallel versions
of PLAST-P  and  the  parallel  version  of the NCBI  BLAST-P program.
Actually, BLAST-P can be run on a parallel machine  by  specifying the
number of nodes. For this experimentation, this option (- a) has  been
set to 2.

\begin{table}[h]
\centering
\small
\begin{tabular}{|c|c|c|c|c|}
\hline  nb. seq.       & multi      & GPU          & multi-GPU     & multicore   \\
           bank2       & PLAST-P    & PLAST-P      & PLAST-P       & BLAST-P     \\
\hline 5k              &   472      &   646        &     379       &    1,953    \\
\hline 10k             &   889      & 1,186        &     692       &    3,794    \\
\hline 20k             & 1,761      & 2,420        &   1,442       &    7,573    \\
\hline 40k             & 3,282      & 4,581        &   2,745       &   14,281    \\
\hline
\end{tabular}
\caption{\small \emph{Execution time (sec) of multicore BLAST-P and various versions of PLAST-P.}}
\label{ComparasionPaPLAST-P}
\end{table}

\begin{table}[h]
\centering
\small
\begin{tabular}{|c|c|c|c|}
\hline  nb. seq.       & multi      & GPU          & multi-GPU    \\
           bank2       & PLAST-P    & PLAST-P      & PLAST-P          \\
\hline 5k              &  4.13      &   3.02       &    5.15          \\
\hline 10k             &  4.26      &   3.19       &    5.48          \\
\hline 20k             &  4.30      &   3.12       &    5.25          \\
\hline 40k             &  4.35      &   3.11       &    5.20          \\
\hline
\end{tabular}
\caption{\small \emph{Speed-up of PLAST-P compared to  multicore BLAST-P}}
\label{SpeedupPLASTPMuitBLAST-P}
\end{table}

Table \ref{ComparasionPaPLAST-P} reports the total execution time for
various size of {\sc bank2}.  Both  the execution times of  multicore
BLAST-P and multicore PLAST-P are reduced by about 55\% when compared
to their   sequential version (Table \ref{PerformancePLASTPBLAST-P}).
Similarly,  the multicore PLAST-P program  achieves  an  acceleration
factor  of 4.2  compared  to  the multicore BLAST-P program.

When  comparing  performances   between   multicore  PLAST-P  and  GPU
PLAST-P,  multicore   PLAST-P  gains   25\%  over  the   GPU   PLAST-P
program. This can be explained as follows: The execution time of stage
3-1 and stage 4 in GPU PLAST-P represents 63\% of the  total execution
time and  only stage  2  and 3-1 are  performed  in  parallel. In  the
multicore GPU PLAST-P version,  the first processor and GPU take about
40\% of the execution time. The second takes about 60\%,  so that this
version  obtained a  40\% speed-up  over  the GPU version.

\subsubsection{Selectivity of stage 2}

We have examined the number  of alignments  found in different PLAST-P
parallel  implementations.   The  percentage  of   ungapped alignments
considered     as     successful     is     presented     in     Table
\ref{ComparasionNbPaPLAST}.   The  percentage  of successful  ungapped
alignments  in multicore   PLAST-P is  the same  as in  the sequential
version (with SIMD optimization). Thus, the number of alignments found
by both versions is also equal.

However, the percentage of successful  ungapped alignments in  the two
GPU versions is higher than the  others  because  of  the  computation
performed on the  GPU: the length of the subsequences (in the ungapped
extension  stage) must  be a  multiple of  16. Thus,  subsequences are
longer  than the  subsequences  of other  versions.  Consequently, the
percentage of successful ungapped alignments increases.

\begin{table}[h]
\centering
\small
\begin{tabular}{|c|c|c|c||c|c|c|}
\hline   nb. seq.  & \multicolumn{3}{|c||}{\% successful ungapped extension}& \multicolumn{3}{|c|}{number of alignments found}\\
\cline{2-7} bank2     & multi      & GPU       & multi-GPU    &  multi        & GPU         & multi-GPU \\
\hline 5k              & 0.186\%   & 0.194\%  & 0.194\%       & 305,502       &  305,663    &  305,662  \\
\hline 10k             & 0.185\%   & 0.193\%  & 0.193\%       & 611,154       &  610,993    &  610,989  \\
\hline 20k             & 0.185\%   & 0.193\%  & 0.193\%       &1,059,796      &1,062,130    & 1,062,129 \\
\hline 40k             & 0.185\%   & 0.192\%  & 0.193\%       &2,237,086      &2,237,140    & 2,237,137 \\
\hline
\end{tabular}
\caption{\small \emph{Results of different PLAST-P parallel implementations.}}
\label{ComparasionNbPaPLAST}
\end{table}

\subsubsection{GPU Execution time}

The  computing  times  of  ungapped and  small gapped  alignments  are
reported in Table \ref{ComparasionUGPLASTP}. The  second column  shows
execution  time of  ungapped  alignment in PLAST-P program (sequential
version  with  filter  optimization).  In  GPU  PLAST-P  program,  GPU
achieved a  speed-up  ranging  from 8.5 to 10. The  execution times of
small gapped alignments are presented in the three last columns. As it
can be seen, GPU also  achieved  a speed-up  ranging  from 9 to 11 for
when comparing with the version without optimization.

\begin{table}[h]
\centering
\small
\begin{tabular}{|c|c|c||c|c|c|}
\hline nb. seq.& \multicolumn{2}{|c||}{ungapped extension}&\multicolumn{3}{|c|}{small gapped extension}\\
\cline{2-6}bank2& filter opt.&     GPU          & no opt.  & SIMD 8-bit    &  GPU        \\
\hline 5k       &  1,511     &     148  (10.21) &    626   &  226   (2.76) &  64  (9.78) \\
\hline 10k      &  2,643     &     295  (8.95)  &   1,232  &  443   (2.78) &  136 (9.05) \\
\hline 20k      &  5,102     &     594  (8.59)  &   2,602  &  822   (3.16) &  275 (9.64) \\
\hline 40k      & 10,256     &   1,079  (9.51)  &   5,213  & 1,564  (3.33) &  470 (11.09)\\
\hline
\end{tabular}
\caption{\small \emph{Execution time (sec) of ungapped and small gapped alignment in
                       different implementations. Numbers in brackets represent speed-up.}}
\label{ComparasionUGPLASTP}
\end{table}

\newpage
\section{TPLAST-N}

A TBLAST-N  search compares a  protein sequence to the  six translated
frames of  a nucleotide database. It  can be a very  productive way of
finding  homologous proteins  in an  unannotated  nucleotide sequence.
Like  BLAST-P,  TBLAST-N is  not  specifically  adapted for  intensive
sequence comparison.  Thus, similarly to PLAST-P,  TPLAST-N uses the
double  indexing  technique  for  speeding  up the  search.   In  this
section, details of implementation are not described. We only focus on
the TPLAST-N program performances.\\

\noindent
All the measures have been done with the following data set:

\begin{itemize}

\item DNA  bank: {\sc bank3} (cf. Annex 1)

\item 4  protein banks:  {\sc bank2} (cf. Annex 1)

\end{itemize}

\subsection{TPLAST-N sequential version}

This section presents the performances of three sequential versions:

\begin{itemize}

\item no optimization: TPLAST-N (no opt.)
\item filter optimization: TPLAST-N (filter)
\item SIMD optimisation: TPLAST-N (SIMD)

\end{itemize}

\subsubsection{TPLAST-N Profiling}

Table \ref{AverageRuntimeTPLASTN} shows the percentage of time  spent
in  the  different  stages of the TPLAST-N (no opt.)  program. Again,
stages 2 and 3.1 are good candidates for parallelization  since   72\%
and 23\% of the execution time is spent in these two stages.

\begin{table}[h]
\centering\small
\begin{tabular}{|c|l|c|}
\hline Stage & Task                                         & Percentage of overall time\\
\hline 0     & Index the two banks                          &  0.4\% \\
\hline 1     & Construct neighborhood blocks                &  0.6\% \\
\hline 2     & Perform ungapped alignment                   & 72.0\% \\
\hline 3-1   & Perform small gapped alignment               & 23.0\% \\
\hline 3-2   & Perform full gapped alignment                &  2.7\% \\
\hline 4     & Trace-Back information and display alignments &  1.3\% \\
\hline
\end{tabular}
\caption{\small \emph{Average runtime for each stage of TPLAST-N (no opt.)}}
\label{AverageRuntimeTPLASTN}
\end{table}

\subsubsection{Execution time}

Table  \ref{PerformanceTPLASTN} shows  the total  execution  time (in
second).   Note that  TPLAST-N  (no opt.) is slower  than
TBLAST-N (about 10 \% slower).

\begin{table}[h]
\centering
\small
\begin{tabular}{|c|c|c|c|c|c|}
\hline nb. seq.    & TPLAST-N     &  TPLAST-N       & TPLAST-N     &  TPLAST-N   & \multirow{2}{*}{TBLAST-N}\\
        bank2      & (no opt.)     &  (filter)        & (SIMD 16-bit) &  (SIMD 8-bit)&          \\
\hline 5k          &  7,817	       &  5,392	          &  1,491        &   1,328      &  7,036   \\
\hline 10k         & 15,261	       &  9,704  	      &  2,789	      &   2,394      & 13,784   \\
\hline 20k         & 30,332        & 18,069           &  5,411        &   4,554      & 27,147   \\
\hline 40k         & 57,987        & 33,540           & 10,232        &   8,553      & 52,232   \\
\hline
\end{tabular}
\caption{\small \emph{Execution time (sec) of TBLAST-N and the three sequential implementations of TPLAST-N.}}
\label{PerformanceTPLASTN}
\end{table}

\begin{table}[h]
\centering
\small
\begin{tabular}{|c|c|c|c|c|}
\hline nb. seq.    & TPLAST-N     &  TPLAST-N       & TPLAST-N      &  TPLAST-N      \\
        bank2      & (no opt.)    &  (filter)       & (SIMD 16-bit) &  (SIMD 8-bit)  \\
\hline 5k          &  0.90	      &  1.30	        &  4.71         &   5.29         \\
\hline 10k         &  0.90	      &  1.42  	        &  4.94	        &   5.75         \\
\hline 20k         &  0.89        &  1.50           &  4.98         &   5.96         \\
\hline 40k         &  0.90        &  1.55           &  5.10         &   6.10         \\
\hline
\end{tabular}
\caption{\small \emph{Speed-up of TPLAST-N compared to TBLAST-N.}}
\label{SpeedupTPLASTNTBLAST-N}
\end{table}

Table \ref{SpeedupTPLASTNTBLAST-N} shows that the 8-bit SIMD implementation
provides  the best performances  with a speed-up  ranging  from 5 to 6
compared to the NCBI TBLAST-N implementation.

Table  \ref{ComparasionUGTPLASTN}  shows the  execution  time (in second) of  the
ungapped alignment stage only (stage 2).

\begin{table}[h]
\centering
\small
\begin{tabular}{|c|c|c|c|c|}
\hline nb. seq.        &  TPLAST-N & TPLAST-N         & TPLAST-N     & TPLAST-N \\
	bank2              &  (no opt.) & (filter)        &(SIMD 16-bit) &(SIMD 8-bit)   \\
\hline 5k              &   6,495    &  4,333          &     832      &    681    \\
\hline 10k             &  12,590    &  7,264          &   1,549      &  1,165    \\
\hline 20k             &  25,385    & 13,325          &   3,032      &  2,173    \\
\hline 40k             &  48,408    & 24,251          &   5,682      &  3,963    \\
\hline
\end{tabular}
\caption{\small \emph{Execution time (sec) of ungapped alignment in different implementations.}}
\label{ComparasionUGTPLASTN}
\end{table}

\begin{table}[h]
\centering
\small
\begin{tabular}{|c|c|c|c|}
\hline nb. seq.        & TPLAST-N       & TPLAST-N       & TPLAST-N     \\
	bank2              & (filter)       & (SIMD 16-bit)  & (SIMD 8-bit) \\
\hline 5k              & 1.49           &     7.80       &   9.53       \\
\hline 10k             & 1.73           &     8.12       &   10.80      \\
\hline 20k             & 1.90           &     8.37       &   11.68      \\
\hline 40k             & 1.99           &     8.50       &   12.21      \\
\hline
\end{tabular}
\caption{\small \emph{Speed-up of ungapped alignment compared to no optimization.}}
\label{SpeedupUGTPLAST-N}
\end{table}

\subsubsection{Selectivity of stage 2}

Table \ref{ComparasionNbTPLASTN}  gives the percentage  of hits which
have produced  successful ungapped alignments (stage 2) and the  total
number of alignments generated by the TPLAST-N programs.

\begin{table}[h]
\centering
\small
\begin{tabular}{|c|c|c|c||c|c|c|}
\hline   nb. seq.  & \multicolumn{3}{|c||}{\% successful ungapped extension}& \multicolumn{3}{|c|}{number of alignments found}\\
\cline{2-7}bank2       & (no opt.)  & (filter)    & (SIMD)     & (no opt.)  & (filter)    &  (SIMD)    \\
\hline 5k              & 0.145\%    & 0.127\%     & 0.146\%    &  290,378   &  290,277    &   290,387  \\
\hline 10k             & 0.148\%    & 0.129\%     & 0.149\%    &  573,238   &  572,057    &   573,248  \\
\hline 20k             & 0.145\%    & 0.127\%     & 0.145\%    &1,173,396   &1,166,243    & 1,172,422  \\
\hline 40k             & 0.147\%    & 0.128\%     & 0.147\%    &2,211,782   &2,198,534    & 2,211,808  \\
\hline
\end{tabular}
\caption{\small \emph{Selectivity of stage 2 and number of alignments found by the TPLAST-N sequential implementations.}}
\label{ComparasionNbTPLASTN}
\end{table}

\subsubsection{Sensibility}

The table \ref{SensitivityTPLASTN} compares the sensitivity of TBLAST-N
and TPLAST-N (SIMD) following the criteria described in Annex 2.

\begin{table}[h]
\centering
\small
\begin{tabular}{|c|c|c|c|c|c|}
\hline   nb. seq          & nb. alig. TBLAST-N & nb. alig. TPLAST-N & 2\%    &    5\% &   10\% \\
\hline 5k                 &   290,016          &   290,387           & 95.6\% & 95.7\% & 95.9\% \\
\hline 10k                &   572,608          &   573,248           & 96.0\% & 96.2\% & 96.4\% \\
\hline 20k                & 1,172,466          & 1,172,422           & 96.4\% & 96.5\% & 96.6\% \\
\hline 40k                & 2,208,330          & 2,211,808           & 96.4\% & 96.6\% & 96.7\% \\
\hline
\end{tabular}
\caption{\small \emph{Comparison of sensitivity between TPLAST-N and TBLAST-N}}
\label{SensitivityTPLASTN}
\end{table}

\subsection{TPLAST-N parallel version}

This section presents the performances of three parallel versions of TPLAST-N:

\begin{itemize}

\item multicore: TPLAST-N (multi)
\item grapics board: TPLAST-N (GPU)
\item multicore + graphics board:  TPLAST-N (multi-GPU)

\end{itemize}

\subsubsection{Execution time}

Table \ref{ComparasionPaTBLASTN} compares the total execution time  of
the parallel TPLAST-N programs and the NCBI TBLAST-N program run with
the -a option.

\begin{table}[h]
\centering
\small
\begin{tabular}{|c|c|c|c|c|}
\hline  nb. seq.       & multi     & GPU          &  multi-GPU    & multicore   \\
           bank2       &  TPLAST-N &   TPLAST-N   & TPLAST-N      &  TBLAST-N   \\
\hline 5k              &   746     &   773        &     617       &    3,859    \\
\hline 10k             & 1,341     & 1,372        &   1,096       &    7,509    \\
\hline 20k             & 2,503     & 2,613        &   2,064       &   14,831    \\
\hline 40k             & 4,741     & 4,942        &   4,016       &   28,674    \\
\hline
\end{tabular}
\caption{\small \emph{Execution time (sec) of multicore TBLAST-N and the three parallel implementations of TPLAST-N}}
\label{ComparasionPaTBLASTN}
\end{table}

\begin{table}[h]
\centering
\small
\begin{tabular}{|c|c|c|c|}
\hline  nb. seq.       & multi      & GPU          & multi-GPU        \\
           bank2       & TPLAST-N   & TPLAST-N     & TPLAST-N         \\
\hline 5k              &  5.17      &   4.99       &    6.25          \\
\hline 10k             &  5.59      &   5.47       &    6.85          \\
\hline 20k             &  5.92      &   5.67       &    7.18          \\
\hline 40k             &  6.04      &   5.80       &    7.13          \\
\hline
\end{tabular}
\caption{\small \emph{Speed-up of TPLAST-N compared to  multicore TBLAST-N}}
\label{SpeedupTPLASTNMuitTBLAST-N}
\end{table}

The  best  performances  are  provided by  the  TPLAST-N  (multi-GPU)
program with  a speed-up  ranging from  6 to 7  compared to the  NCBI
TBLAST-N program as showed in Table \ref{SpeedupTPLASTNMuitTBLAST-N}.

Table  \ref{ComparasionUGTPLASTNTOTAL} indicates  the  sequential and
parallel execution  times of stage 2 (ungapped  alignments) and stage
3.1 (small gapped alignments).

\begin{table}[h]
\centering
\small
\begin{tabular}{|c|c|c||c|c|c|}
\hline nb. seq. & \multicolumn{2}{|c||}{ungapped extension}&\multicolumn{3}{|c|}{small gapped extension}\\
\cline{2-6}bank2&  filter     & GPU             & no opt.  & SIMD 8-bit   &  GPU         \\
\hline 5k       &  4,333      &     378 (11.46) &   1,050  &   436 (2.40) &  120  (8.75) \\
\hline 10k      &  7,264      &     703 (10.33) &   2,103  &   872 (2.41) &  259  (8.11) \\
\hline 20k      & 13,325      &   1,361  (9.79) &   4,088  & 1,686 (2.42) &  462  (8.48) \\
\hline 40k      & 24,251      &   2,589  (9.36) &   7,866  & 3,233 (2.43) &  893  (8.80) \\
\hline
\end{tabular}
\caption{\small \emph{Execution time (sec) of ungapped and small gapped alignment in different
                      implementations. Numbers in brackets represent speed-up.}}
\label{ComparasionUGTPLASTNTOTAL}
\end{table}

\subsubsection{Selectivity of stage 2}

Table \ref{ComparasionNbPaTPLASTN} gives the percentage of hits which
have produced  successful ungapped alignments (stage 2) and the  total
number of alignments generated by the parallel versions of TPLAST-N.

\begin{table}[h]
\centering
\small
\begin{tabular}{|c|c|c|c||c|c|c|}
\hline   nb. seq.  & \multicolumn{3}{|c||}{\% successful ungapped extension}& \multicolumn{3}{|c|}{number of alignment found}\\
\cline{2-7} bank2     & multi      & GPU       & multi-GPU    &  multi         & GPU         & multi-GPU \\
\hline 5k              & 0.146\%   & 0.154\%  & 0.154\%       &  290,378       &  290,406    &  290,403  \\
\hline 10k             & 0.149\%   & 0.158\%  & 0.158\%       &  573,252       &  573,880    &  573,881  \\
\hline 20k             & 0.145\%   & 0.153\%  & 0.153\%       &1,172,415       &1,173,378    &1,173,378  \\
\hline 40k             & 0.147\%   & 0.156\%  & 0.156\%       &2,211,867       &2,211,935    &2,211,953  \\
\hline
\end{tabular}
\caption{\small \emph{Selectivity of stage 2 and number of alignments found by the TPLAST-N parallel implementations.}}
\label{ComparasionNbPaTPLASTN}
\end{table}

\newpage
\section{PLAST-X}

BLAST-X  compares  translational  products  of  the  nucleotide  query
sequence  to a  protein database. BLAST-X  is often the first analysis
performed  with a  newly determined  nucleotide sequence. Like BLAST-P
and  TBLAST-N,  BLAST-X  is not  specifically  adapted  for  intensive
sequence  comparison. Thus, similarity to PLAST-P,  PLAST-X uses the
double indexing  technique for speeding up the search. In this section,
details of  implementation  are  not  described. We  only focus on the
PLAST-X program performances.\\

\noindent
All the measures have been done with the following data set:

\begin{itemize}

\item DNA bank: {\sc bank3} (cf. Annex 1)

\item 4 DNA banks:  {\sc bank4} (cf. Annex 1)

\end{itemize}

\subsection{PLAST-X sequential version}

This section presents the performances of three sequential versions:

\begin{itemize}

\item no optimization: PLAST-X (no opt.)
\item filter optimization: PLAST-X (filter)
\item SIMD optimisation: PLAST-X (SIMD)

\end{itemize}

\subsubsection{PLAST-X Profiling}

Table \ref{AverageRuntimePLASTX} shows the  percentage of  time spent
in the  different  stages  of  the  PLAST-X (no opt.) program. Again,
stages 2 and 3.1 are  good  candidates for parallelization since 79\%
and 13.8\% of the execution time is spent in these two stages.

\begin{table}[h]
\centering\small
\begin{tabular}{|c|l|c|}
\hline Stage & Task                                         & Percentage of overall time\\
\hline 0     & Index the two banks                      &  0.4\% \\
\hline 1     & Construct neighborhood blocks                &  0.5\% \\
\hline 2     & Perform ungapped alignment                   & 79.0\% \\
\hline 3-1   & Perform small gapped alignment               & 13.8\% \\
\hline 3-2   & Perform full gapped alignment                &  5.3\% \\
\hline 4     & Trace-Back information and display alignments &  1.0\% \\
\hline
\end{tabular}
\caption{\small \emph{Average runtime for each stage of PLAST-X (no opt.)}}
\label{AverageRuntimePLASTX}
\end{table}

\subsubsection{Execution time}

Table  \ref{PerformancePLASTX}  shows  the  total execution  time (in
second). Note that  PLAST-X (no opt.) is slower  than  BLAST-X (about
20 \% slower).

\begin{table}[h]
\centering
\small
\begin{tabular}{|c|c|c|c|c|c|}
\hline nb. seq.    & PLAST-X     &PLAST-X         & PLAST-X      & PLAST-X     & \multirow{2}{*}{BLAST-X}\\
        bank4      & (no opt.)    & (filter)        & (SIMD 16-bit) &( SIMD 8-bit) &          \\
\hline 1k          & 1,775	      & 1,210	        &    369 	    &     328      &  1,501   \\
\hline 3k          & 5,583	      & 3,345  	        &  1,186	    &   1,028      &  4,455   \\
\hline 6k          &10,737        & 6,262           &  2,302        &   1,989      &  8,495   \\
\hline 10k         &18,125        &10,588           &  4,089        &   3,523      & 14,035   \\
\hline
\end{tabular}
\caption{\small \emph{Performance between PLAST-X and BLAST-X.}}
\label{PerformancePLASTX}
\end{table}

\begin{table}[h]
\centering
\small
\begin{tabular}{|c|c|c|c|c|}
\hline nb. seq.    & PLAST-X      &  PLAST-X        & PLAST-X       &  PLAST-X      \\
        bank2      & (no opt.)    &  (filter)       & (SIMD 16-bit) &  (SIMD 8-bit)  \\
\hline 1k          &  0.84	      &  1.24	        &  4.06         &   4.57         \\
\hline 3k          &  0.79	      &  1.33  	        &  3.76	        &   4.33         \\
\hline 6k          &  0.79        &  1.35           &  3.69         &   4.27         \\
\hline 10k         &  0.77        &  1.32           &  3.43         &   3.98         \\
\hline
\end{tabular}
\caption{\small \emph{Speed-up of PLAST-X compared to BLAST-X.}}
\label{SpeedupPLASTXBLAST-X}
\end{table}

The 8-bit SIMD implementation provides  the best  performances  with a
speed-up  ranging  from  4  to 4.5  compared  to  the   NCBI   BLAST-X
implementation.

Table \ref{ComparasionUGPLASTX} shows the  execution time (in second)
of the ungapped alignment stage only (stage 2).

\begin{table}[h]
\centering
\small
\begin{tabular}{|c|c|c|c|c|}
\hline nb. seq.        &  PLAST-X & PLAST-X    & PLAST-X      & PLAST-X    \\
        bank4          & (no opt.) & (filter)    &(SIMD 16-bit)  &(SIMD 8-bit) \\
\hline 1k              & 1,449     &   918       &     182       &    143      \\
\hline 3k              & 4,424     & 2,338       &     530       &    378      \\
\hline 6k              & 8,439     & 4,312       &   1,004       &    700      \\
\hline 10k             &13,941     & 6,994       &   1,680       &  1,156      \\
\hline
\end{tabular}
\caption{\small \emph{Execution time (sec) of ungapped alignment in different implementations.}}
\label{ComparasionUGPLASTX}
\end{table}

\begin{table}[h]
\centering
\small
\begin{tabular}{|c|c|c|c|}
\hline nb. seq.        & PLAST-X        & PLAST-X        & PLAST-X     \\
	bank2              & (filter)       & (SIMD 16-bit)  & (SIMD 8-bit) \\
\hline 1k              & 1.57           &     7.96       &   10.13       \\
\hline 3k              & 1.89           &     8.34       &   11.70      \\
\hline 6k              & 1.95           &     8.40       &   12.05      \\
\hline 10k             & 1.99           &     8.29       &   12.05      \\
\hline
\end{tabular}
\caption{\small \emph{Speed-up of ungapped alignment compared to no optimization.}}
\label{SpeedupUGPLAST-X}
\end{table}

\subsubsection{Selectivity of stage 2}

Table \ref{ComparasionNbPLASTX} gives the  percentage  of hits which
have produced successful ungapped alignments (stage 2) and the  total
number of alignments generated by the PLAST-X programs.

\begin{table}[h]
\centering
\small
\begin{tabular}{|c|c|c|c||c|c|c|}
\hline nb. seq.  & \multicolumn{3}{|c||}{\% successful ungapped extension }& \multicolumn{3}{|c|}{number of alignments found}\\
\cline{2-7}bank4       & (no opt.)  & (filter)    & (SIMD)     & (no opt.)  & (filter)    &  (SIMD)   \\
\hline 1k              & 0.150\%    & 0.132\%     & 0.150\%    &  32,326    &  32,310     &   32,328  \\
\hline 3k              & 0.152\%    & 0.134\%     & 0.152\%    &  97,059    &  97,010     &   97,059  \\
\hline 6k              & 0.154\%    & 0.135\%     & 0.154\%    & 196,245    & 196,168     &  196,252  \\
\hline 10k             & 0.159\%    & 0.140\%     & 0.159\%    & 332,124    & 331,928     &  332,135  \\
\hline
\end{tabular}
\caption{\small \emph{Selectivity of stage 2 and number of alignments found by the PLAST-X sequential implementations.}}
\label{ComparasionNbPLASTX}
\end{table}

\subsubsection{Sensibility}

The table \ref{SensitivityPLASTX} compares the sensitivity of BLAST-X and
PLAST-X (SIMD) following the criteria of the Annex 2.

\begin{table}[h]
\centering
\small
\begin{tabular}{|c|c|c|c|c|c|}
\hline   nb. seq & nb. alig. BLAST-X& nb. alig. PLAST-X     &   2\%   &  5\%    &   10\% \\
\hline 1k                 & 31,873        &   32,328         & 94.4\%  & 97.8\%  & 98.3\% \\
\hline 3k                 & 94,221        &   97,059         & 94.7\%  & 97.9\%  & 98.3\% \\
\hline 6k                 &189,488        &  196,252         & 94.6\%  & 98.0\%  & 98.5\% \\
\hline 10k                &321,416        &  332,135         & 94.6\%  & 98.0\%  & 98.5\% \\
\hline
\end{tabular}
\caption{\small \emph{Comparison of sensitivity between PLAST-X and BLAST-X}}
\label{SensitivityPLASTX}
\end{table}

\subsection{PLAST-X parallel version}

This section presents the performances of three parallel versions of PLAST-X:

\begin{itemize}

\item multicore: PLAST-X (multi)
\item grapics board: PLAST-X (GPU)
\item multicore + graphics board:  PLAST-X (multi-GPU)

\end{itemize}

\subsubsection{Execution time}

Table \ref{ComparasionPaBLASTX}  compares the  total execution time of
the parallel PLAST-X programs  and the NCBI BLAST-X  program run with
the -a option.

\begin{table}[h]
\centering
\small
\begin{tabular}{|c|c|c|c|c|}
\hline  nb. seq.     & multi      &   GPU        & multi-GPU    & multicore  \\
      bank4          &  PLAST-X   & PLAST-X      &  PLAST-X     & BLAST-X    \\
\hline  1k           &    178     &     191      &     129      &      833   \\
\hline  3k           &    564     &     662      &     379      &    2,403   \\
\hline  6k           &  1,093     &   1,309      &     778      &    4,622   \\
\hline 10k           &  1,937     &   2,462      &   1,531      &    7,550   \\
\hline
\end{tabular}
\caption{\small \emph{Execution time (sec) of multicore BLAST-X and the three parallel implementations of PLAST-X.}}
\label{ComparasionPaBLASTX}
\end{table}

\begin{table}[h]
\centering
\small
\begin{tabular}{|c|c|c|c|}
\hline  nb. seq.       & multi      & GPU          & multi-GPU        \\
           bank2       & PLAST-X    & PLAST-X      & PLAST-X          \\
\hline 1k              &  4.60      &   4.36       &    6.45          \\
\hline 3k              &  4.26      &   3.62       &    6.34          \\
\hline 6k              &  4.22      &   3.53       &    5.94          \\
\hline 10k             &  3.89      &   3.06       &    4.93          \\
\hline
\end{tabular}
\caption{\small \emph{Speed-up of PLAST-X compared to  multicore BLAST-X}}
\label{SpeedupPLASTXMuitBLAST-X}
\end{table}

The  best  performances  are  provided by  the  PLAST-X  (multi-GPU)
program with a speed-up ranging from  5 to 6 compared to  the  NCBI
BLAST-X program.

Table  \ref{ComparasionUGPLASTXTOTAL} indicates  the  sequential  and
parallel execution  times of stage 2 (ungapped  alignments) and stage
3.1 (small gapped alignments).

\begin{table}[h]
\centering
\small
\begin{tabular}{|c|c|c||c|c|c|}
\hline nb. seq.& \multicolumn{2}{|c||}{ungapped extension}&\multicolumn{3}{|c|}{small gapped extension}\\
\cline{2-6}bank4&  filter opt.& GPU         & no opt.  & SIMD 8-bit      &  GPU  \\
\hline 1k       &  918        &  81 (11.33) &    241   &   101 (2.38)    &  30 (8.03)  \\
\hline 3k       &2,338        & 246 (9.50)  &    762   &   319 (2.38)    &  94 (8.10)  \\
\hline 6k       &4,312        & 474 (9.09)  &   1478   &   612 (2.41)    & 166 (8.90)  \\
\hline 10k      &6,998        & 854 (8.19)  &   2568   & 1,064 (2.41)    & 316 (8.12)  \\
\hline
\end{tabular}
\caption{\small \emph{Execution time (sec) of ungapped and small gapped alignment
                      in different implementations. Numbers in brackets represent speed-up.}}
\label{ComparasionUGPLASTXTOTAL}
\end{table}

\subsubsection{Selectivity of stage 2}

Table \ref{ComparasionNbPaPLASTX}  gives the percentage  of hits which
have produced  successful ungapped alignments (stage 2) and the  total
number of alignments generated by the parallel versions of PLAST-X.

\begin{table}[h]
\centering
\small
\begin{tabular}{|c|c|c|c||c|c|c|}
\hline   nb. seq.  & \multicolumn{3}{|c||}{\% successful ungapped extension}& \multicolumn{3}{|c|}{number of alignments found}\\
\cline{2-7} bank4      &   multi   & GPU      & multi-GPU     &  multi        & GPU        & multi-GPU \\
\hline 1k              & 0.150\%   & 0.158\%  & 0.158\%       &  32,330       &  32,344    &  32,344   \\
\hline 3k              & 0.153\%   & 0.161\%  & 0.161\%       &  97,059       &  97,173    &  97,173   \\
\hline 6k              & 0.154\%   & 0.162\%  & 0.162\%       & 196,256       & 196,514    & 196,514   \\
\hline 10k             & 0.160\%   & 0.168\%  & 0.168\%       & 332,133       & 332,418    & 332,418   \\
\hline
\end{tabular}
\caption{\small \emph{Selectivity of stage 2 and number of alignments found by the PLAST-X parallel implementations}}
\label{ComparasionNbPaPLASTX}
\end{table}

\newpage
\section{Comparison with the BLAST-P family}

This section summarizes the performances of the various PLAST-P implementations
by comparing the speed-up between the PLAST-P and BLAST-P families.

\subsection{BLAST-P vs PLAST-P}

\begin{figure}[h]
  \centering
  \includegraphics[width=5in]{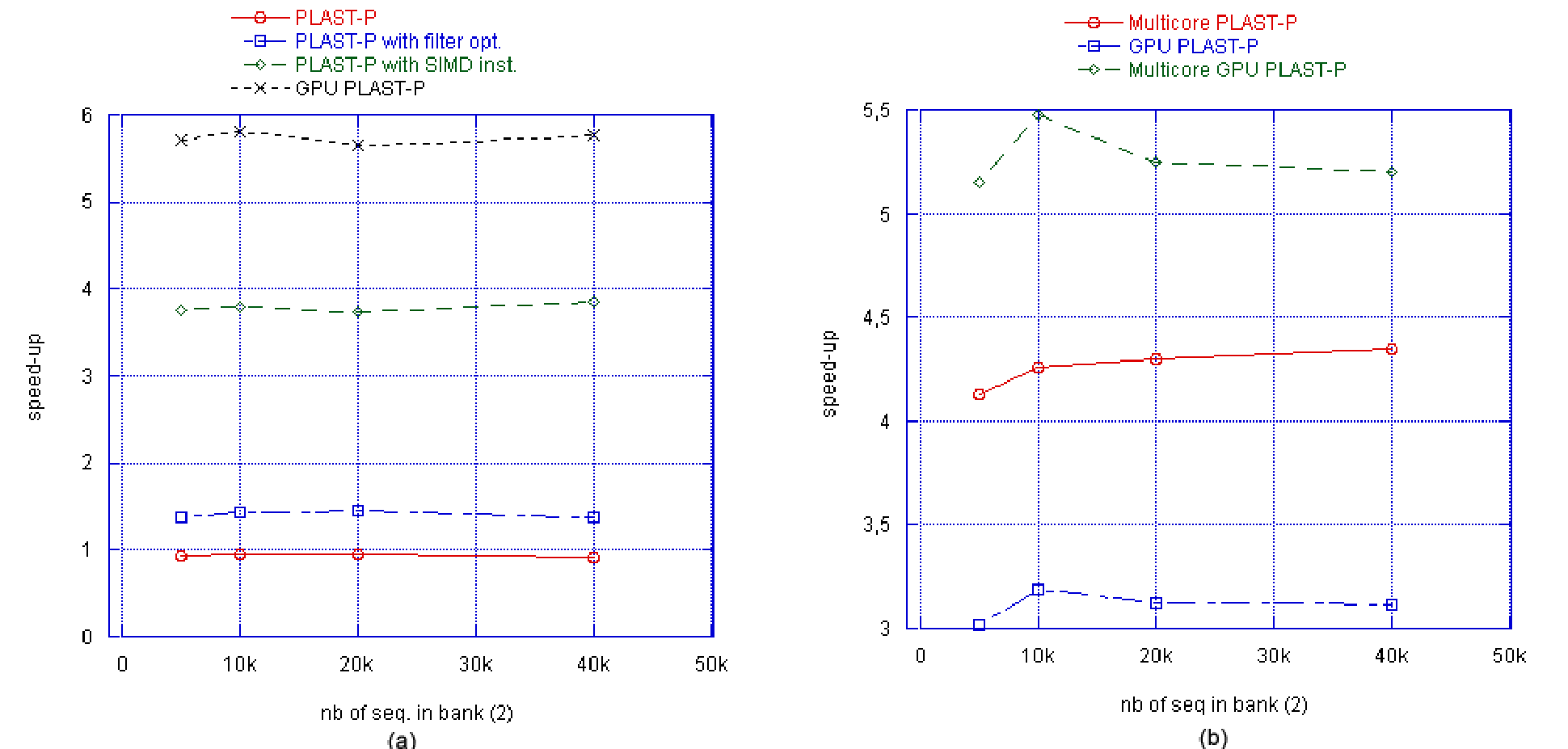}
  \caption{\small \emph{Speed-up of different PLAST-P: (a) over BLAST-P; (b) over multicore BLAST-P.}}
\label{Fig:TBLASTPNSP}
\end{figure}

\subsection{TBLAST-N vs TPLAST-N}

\begin{figure}[h]
  \centering
  \includegraphics[width=5in]{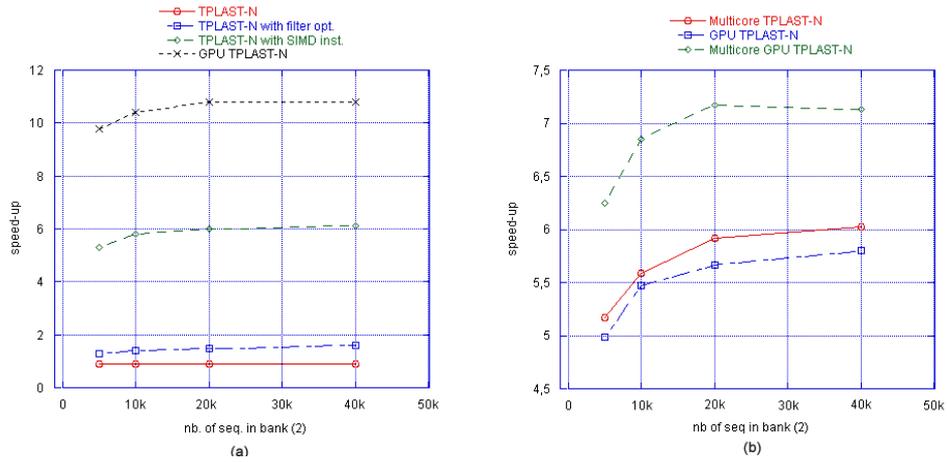}
  \caption{\small \emph{Speed-up of different TPLAST-N: (a) over TBLAST-N; (b) over multicore TBLAST-N.}}
\label{Fig:TBLASTNSP}
\end{figure}

\subsection{BLAST-X vs PLAST-X}

\begin{figure}[h]
  \centering
  \includegraphics[width=5in]{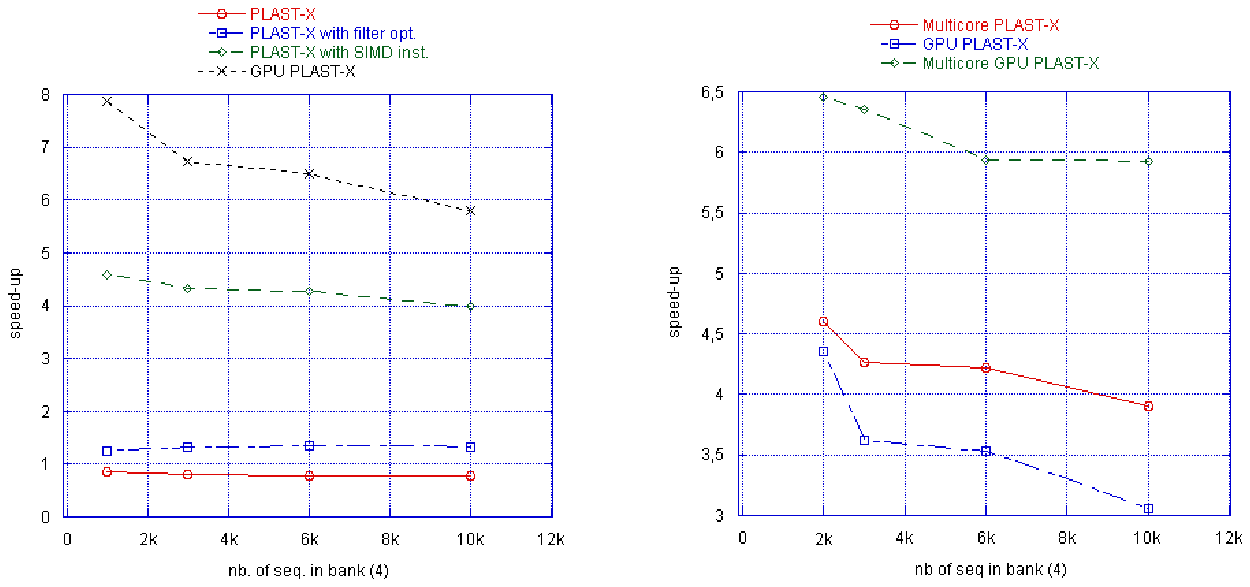}
  \caption{\small \emph{Speed-up of different PLAST-X: (a) over BLAST-X; (b) over multicore BLAST-X.}}
\label{Fig:BLASXSP}
\end{figure}

\newpage

\section*{Annex 1: Hardware and Data Set}

\subsection*{Hardware Platform}

\subsubsection*{Processor}

We tested the experiments on: An Intel core 2 Dual 2.6 GH processor with
2 MB L2 cache, and 2 GB RAM, running  Linux (fedora 6).

\subsubsection*{Graphic board}

We used the graphic card GeForce 8800 GTX
(version GPU). The characteristics of this board are the following:

\begin{itemize}
 \item 16 multiprocessors SIMD at 675 MHz; each multiprocessor is
       composed of eight processors running at twice the clock frequencies;
 \item the maximum number of threads per block is 512;
 \item the amount of device memory is 768 MB at 1.8 GHz engine clock speed;
 \item the maximum bandwidth observed between the computer memory and the device
       memory is 2 GB/s.
\end{itemize}

\subsection*{Data set}

\begin{itemize}

\item {\sc bank1}: It contents  141,708  sequences from the PIR-Protein  bank
(version 80 01/2005) with  an average length of 340;

\item {\sc bank2}: Actually, this is a set of four banks extracted from the  SWISS-PROT bank (version 05/2007)
which content respectively 5,000; 10,000; 20,000, and 40,000 sequences with an average length of 367;

\item {\sc bank3}: It contents 27,360 sequences (gbvrt3 in GenBank,  version 156)  with  an
average  length  of 5,454;

\item {\sc bank4}: This is a set of four banks extracted from the gbvrl division of Genbank (version 156)
which content respectively 1,000; 3,000; 6,000, and 10,000 sequences with an average length of 1,024.

\end{itemize}

\newpage
\section*{Annex 2: Measure criteria}

\subsection*{Execution time}

The execution time is calculated using the Linux time command. For each experiment, the machine
is only dedicated to the computation under test. BLAST was launched with the default settings,
except for the e-value statistical parameter set to $10^{-3}$. It represents
a reasonable value in the context of intensive sequence comparison.

\subsection*{Sensibility}

The sensitivity is evaluated in relation to the number of
alignments found in both cases. Specifically, we look at whether alignments found
begin and end at the same places in the two sequences with a margin of Y~\% calculated
on the average size of the 2 alignments. For example, to compare two alignments of size
100 with 5\% margin, we check that the start and end positions for making
up this alignment are within a range of 5 amino acids.

\newpage

\section*{Annex 3: Source codes of SIMD ungapped alignment}

\small
\begin{verbatim}
    VectUngappedExtend(char **block1, char **block2,
                                   int n1, int n2, HIT *listeRS){

       // n1, n2: number of subsequences in block1 and block2
       short *databk;
       short *pc;
       short score_arr[];
       __m128i	*pvb, pvScore, vscore, vMaxScore;
       // declaration of the sequence profile memory		
   	   databk = (short *) calloc(SIZE_MT*LenSubSeq, sizeof(__m128i));
       pvb = (__m128i *) databk;
       for(j=0;j<n1;j+=SIZEV){ // SIZEV: number of elements in vector
         pc = (short *) pvb;
         // initiation profile for SIZEV subsequences
         for(i=0;i<SIZE_MT;i++){ // SIZE_MT: size of substitution matrix
            matrixRow = MATRIX[i];  // MATRIX: substitution matrix
            for(k=0;k<LenSubSeq;k++){
               for(l=0;l<SIZEV;l++){
                  *pc++ = (short) matrixRow[block1[j+l][k]];
               }
            }
         }
         // compute scores between SIZEV subsequences in bloc1
         // and all sequences in block2
         for(i=0;i<n2;i++){
            vMaxScore = _mm_xor_si128 (vMaxScore, vMaxScore);
            vscore = _mm_xor_si128 (vscore, vscore);
            for(k=0;k<LenSubSeq;k++){
               pvScore = *(pvb + (block2[i][k] * LenSubSeq + k));
               vscore = _mm_adds_epi16 (vscore, pvScore);
               vMaxScore = _mm_max_epi16 (vMaxScore, vscore);
            }
            // extraction element score
            for(i=0;i<SIZEV;i++)
               score_arr[i] =  _mm_extract_epi16(vMaxScore,i);
            for(k=0;k<SIZEV;k++)
              if(score_arr[k]>=S2) add(ungapped, listRS);
          }
       }
    }
\end{verbatim}

\newpage
\section*{Annex 4: Source gode of GPU ungapped alignment}

\small
\begin{verbatim}
   GPUngappedExtend(char* h_A, char* h_B, char* h_C, int nba, int nbb){
      // nba, nbb: number of subsequences in block A and block B
      // allocate device memory
      unsigned char* d_A;
      CUDA_SAFE_CALL(cudaMalloc((void**) &d_A, mem_size_A));
      unsigned char* d_B;
      CUDA_SAFE_CALL(cudaMalloc((void**) &d_B, mem_size_B));

      // copy host memory to device
      CUDA_SAFE_CALL(cudaMemcpy(d_A, h_A, mem_size_A,
                              cudaMemcpyHostToDevice) );
      CUDA_SAFE_CALL(cudaMemcpy(d_B, h_B, mem_size_B,
                              cudaMemcpyHostToDevice) );

      // allocate device memory for result
      unsigned int mem_size_C = sizeof(char) * nba * nbb;
      char* d_C;
      CUDA_SAFE_CALL(cudaMalloc((void**) &d_C, mem_size_C));

      // setup execution parameters for block grid
      dim3 threads(BLOCK_SIZE, BLOCK_SIZE);
      dim3 grid(nbb / threads.x, nba / threads.y);

      // execute the kernel
      GPUnGapped_kernel<<< grid, threads >>>(d_C, d_A, d_B, nba, nbb);

      // copy result from device to host
      CUDA_SAFE_CALL(cudaMemcpy(h_C, d_C, mem_size_C,
                              cudaMemcpyDeviceToHost));
   }

   GPUngapped_kernel(char* d_C, char* d_A, char* d_B, int nba, int nbb){

      // block index
      int bx = blockIdx.x;
      int by = blockIdx.y;
      // thread index
      int tx = threadIdx.x;
      int ty = threadIdx.y;
      // index at the beginning of h_A
      int aBegin = nba * BLOCK_SIZE * by;
      aBegin += nba * ty + tx;
      // step size used to iterate through the sub-bloc of A
      int aStep  = BLOCK_SIZE;
      // index at the beginning of h_B
      int bBegin = __mul24(BLOCK_SIZE,bx);
      bBegin += nbb * ty + tx;
      // step size used to iterate through the sub-tabe of h_B
      int bStep  = __mul24(BLOCK_SIZE,nbb);
      // Csub is used to store the elements of the block sub-score
      //that is computed by the thread
      int Csub = 0;
      int CsubMaxi = 0;
      // declaration of the shared memory array As used to
      // store the sub-block of h_A
      __shared__ int As[BLOCK_SIZE][2*BLOCK_SIZE];
      // declaration of the shared memory array Bs used to
      // store the sub-table of h_B
      __shared__ int Bs[2*BLOCK_SIZE][BLOCK_SIZE];
      // load the matrices from device memory to shared memory;
      AS(ty, tx) = A[aBegin];
      BS(ty, tx) = B[bBegin];
      AS(ty, tx + aStep) = A[aBegin + aStep];
      BS(ty + aStep, tx) = B[bBegin + bStep];
      // synchronize to make sure the sub-blocks are loaded
      __syncthreads();
      // calculate the score two sub_blocks together;
      for (int k = 0; k < 2*BLOCK_SIZE; k++){
        Csub = Csub + texfetch(matrix, AS(ty, k), BS(k, tx));
        if(Csub>CsubMaxi) CsubMaxi = Csub;
        if(k==MAX_RIGHT) Csub = CsubMaxi;
      }
      __syncthreads();
      AS(ty, tx) = A[aBegin + 2*aStep];
      BS(ty, tx) = B[bBegin + 2*bStep];
      // synchronize to make sure the sub-blocks are loaded
      __syncthreads();
      for (int k = 0; k < BLOCK_SIZE; k++){
        Csub = Csub + texfetch(matrix, AS(ty, k), BS(k, tx));
        if(Csub>CsubMaxi) CsubMaxi = Csub;
      }
      // write the block sub-matrix to device memory,
      // each thread writes one element
      int c = bStep * by + bBegin;
      C[c] = CsubMaxi;
}
\end{verbatim}

\end{document}